\documentclass[prd,reprint,aps,amsmath,amssymb,showpacs,nofootinbib]{revtex4-1}
\usepackage[utf8x]{inputenc}
\usepackage{graphicx}
\usepackage{enumitem}
\usepackage{slashed}
\usepackage{url}
\usepackage[pdftex]{hyperref}
\usepackage[T1]{fontenc}
\usepackage{bm}
\usepackage{color}

\newcommand{\bra}[1]{\langle #1|}
\newcommand{\ket}[1]{|#1 \rangle}
\newcommand{\calo}{\mathcal{O}} 
\newcommand{\call}{\mathcal{L}} 
\newcommand{\calm}{\mathcal{M}} 

\begin{document}
\title{\emph{CP} Violation in Pseudo-Dirac Fermion Oscillations}
\author{Seyda Ipek}
\email{ipek@uw.edu}
\affiliation{Department of Physics, University of Washington, Seattle, WA 98195, USA}

\author{David McKeen}
\email{dmckeen@uw.edu}
\affiliation{Department of Physics, University of Washington, Seattle, WA 98195, USA}

\author{Ann E. Nelson}
\email{aenelson@uw.edu}
\affiliation{Department of Physics, University of Washington, Seattle, WA 98195, USA}


\begin{abstract}

Supersymmetric theories with a $U(1)_R$ symmetry have Dirac gauginos, solve the supersymmetric flavor and \emph{CP} problems, and have distinctive collider signatures. However when supergravity is included, the $U(1)_R$ must be broken, adding small Majorana mass terms which split the mass of the two components of the Dirac gaugino and lead to   oscillations between  $U(1)_R$ charge eigenstates. We present a general study of fermion-antifermion oscillations in this system, including the effects of decays and \emph{CP} violation. We consider the effects of such oscillations in the case where the two $U(1)_R$ charge eigenstates can decay into the same final state, and show that $O(1)$ \emph{CP} violation is allowed. In the case of decays into final states containing leptons such \emph{CP} violation can be observed as a same sign dilepton asymmetry. 
\end{abstract}

\maketitle
\section{Introduction}
The universe provides clear evidence for \emph{CP} violation beyond the standard model (SM). Assuming cosmological inflation erases any initial asymmetry,  the asymmetry between matter and antimatter must arise due to a nonequilibrium microphysical process called baryogenesis, which requires \emph{CP} and baryon number violation, that creates the observed asymmetry of $10^{-8}$ between quarks and antiquarks. While the SM weak interactions violate baryon number via nonperturbative processes which are fairly rapid at high temperature~\cite{Kuzmin:1985mm},  the effects of SM \emph{CP} violation are suppressed in the early Universe and new sources of \emph{CP} violation are needed to explain baryogenesis~\cite{Huet:1994jb,*Gavela:1994dt}. However, any new sources of \emph{CP} violation are strongly constrained by searches for electric dipole moments (EDMs) (see, e.g.,~\cite{Pospelov:2005pr,*Cirigliano:2006dg,*Fromme:2006cm,*Morrissey:2012db,*McKeen:2012av,*McKeen:2013dma,*Ipek:2013iba,*Ellis:2008zy}).  For example, in the minimal supersymmetric standard model (MSSM), electron and quark EDMs are generated at one loop level. Consequently, unless the superpartners are very heavy, the \emph{CP}-violating phases in the soft supersymmetry-breaking terms of the MSSM Lagrangian are tightly limited by the null results of EDM experiments~\cite{Baron:2013eja,*Griffith:2009zz,*Baker:2006ts}. Addressing this by assuming \emph{CP} conservation in the SUSY-breaking terms is {\it ad hoc} and limits the possibilities for baryogenesis. Similar considerations apply for flavor-changing neutral currents.

One model that circumvents the fine-tuning of \emph{CP}-violating and flavor-changing terms in the SUSY Lagrangian is the $R$-symmetric MSSM~\cite{Hall:1990hq,*Kribs:2007ac,*Dudas:2013gga}. This model is an extension of the MSSM in which there is a global $U(1)_R$ symmetry~\cite{Fayet:1974pd}. The superpartners of fermions and gauge fields have $R$ charges of $+1$. This charge assignment forbids Majorana masses for the gauginos. Therefore, EDMs through neutralino exchange can only be induced at higher than one loop order, greatly reducing the constraints on the \emph{CP}-violating phases in the Lagrangian. To give mass to the gauginos one needs to add an extra adjoint field with \emph{R} charge $-1$ for each of the MSSM gauginos to allow for Dirac masses. Each gaugino is then paired with its partner to form a Dirac spinor~\cite{Fayet:1975yi}. This scenario has a plausible short-distance origin. Mediating supersymmetry breaking to the MSSM by the nonvanishing expectation value of the $D$-term of a hidden $U(1)$ gauge field leads to $U(1)_R$-preserving Dirac gaugino masses~\cite{Dine:1992yw,Fox:2002bu}. If the sector which mediates supersymmetry breaking  does not contain a gauge singlet field with a non-vanishing $F$-term then Majorana gaugino masses and other $U(1)_R$ symmetry breaking soft supersymmetry breaking terms are suppressed.

The $U(1)_R$ symmetry is expected to be only an approximate symmetry in locally supersymmetric theories because it is always broken by the gravitino mass, and anomaly mediation will produce small $U(1)_R$-breaking Majorana mass terms for the gauginos~\cite{Randall:1998uk,*Giudice:1998xp}. These Majorana masses produce a small mass splitting between different linear combinations of particle and antiparticle states (the eigenstates of $U(1)_R$ with opposite eigenvalues) that make up the Dirac spinor. Since the $U(1)_R$ symmetry is only approximate, the gaugino in this setup is therefore called a pseudo-Dirac fermion. The mass splitting causes oscillations between the two \emph{R} charge eigenstates in a way similar to neutral meson oscillations. Oscillations of pseudo-Dirac neutralinos have been considered in~\cite{Grossman:2012nn}. Another example of pseudo-Dirac fermion oscillation that arises in the context of $R$-symmetric SUSY is mesino-antimesino oscillation~\cite{Sarid:1999zx,*Berger:2012mm}. Pseudo-Dirac fermion oscillations have also been extensively considered in the context of neutrinos~\cite{RevModPhys.59.671,*Wolfenstein1981147,*Petcov1982245,*Kobayashi:2000md,*PhysRevD.22.2227}. However, the \emph{CP} violation found in these systems due to oscillations among the three (or more) generations of neutrinos~\cite{Bray:2007ru} is different from the particle-antiparticle oscillations that we will consider.

\emph{CP} violation in particle-antiparticle oscillations can occur if both states decay into common final states. In Ref.~\cite{Grossman:2012nn} \emph{CP} violation in neutralino oscillations was not considered since it was assumed that there were no final states common to both \emph{R} charge eigenstates. However, there can be common final states when one allows for $U(1)_R$ violating interactions for both the neutralino and its Dirac partner.

In this paper we study \emph{CP} violation in pseudo-Dirac fermion oscillations. For a concrete example with distinctive phenomenology  we consider a pseudo-Dirac gluino which is the lightest MSSM superpartner (besides the gravitino), decaying via R-parity violation.  We show that depending on the parameters of the model,  there can be $O(1)$ \emph{CP} violation in the oscillations. We also comment on ways to observe the \emph{CP} violation from these oscillations, e.g. as a same sign dilepton asymmetry.

The organization of this paper is as follows.  In Sec.~\ref{sec:pdfermion} we set up the formalism to describe pseudo-Dirac fermion oscillations. In Sec.~\ref{sec:model} we give the details of our model and compute the \emph{CP} violation from interference between mixing and decay in pseudo-Dirac gluino oscillations in Sec.~\ref{sec:CP}. A set of benchmark parameters of the model that lead to an interesting signal of \emph{CP} violation is given in Sec.~\ref{sec:benchmark}. Section~\ref{sec:conc} contains concluding remarks on this and variant scenarios.

\section{Pseudo-Dirac Fermion Oscillations} \label{sec:pdfermion}
Before considering a specific model we demonstrate how the addition of a small Majorana mass  to a theory with an otherwise Dirac fermion results in particle-antiparticle oscillations  and obtain the Hamiltonian relevant to this two-state system. As \emph{CP} violation in pseudo-Dirac fermion oscillations has not been previously  discussed  in the literature, we give a detailed treatment of  the formalism here. In order to introduce these oscillations we use the two-component Weyl spinor techniques laid out in~\cite{Dreiner:2010}. 

 In a realistic supersymmetric theory, the pseudo-Dirac fermion could be a neutral mesino~\cite{Sarid:1999zx,*Berger:2012mm}, a neutralino as considered in~\cite{Grossman:2012nn}, or a gluino.
  Reference~\cite{Grossman:2012nn} briefly mentions gluino oscillations, but claims that any macroscopic coherent oscillations would be destroyed due to strong interactions with the detector. However, both the gluino and its Dirac partner are color octet fermions with identical strong interactions. Therefore, as explained in the Appendix, the coherence of the two \emph{R} charge states is not affected by scattering due to strong interactions or gluino hadronization, and such oscillations could be observable. For a concrete simple example of pseudo-Dirac fermion oscillations, we will consider the oscillations of a Dirac gluino here.

We start with a pair of left-handed, color octet Weyl spinors, $\lambda_\alpha$ and $\calo_\alpha$ where $\alpha=1,2$ is the spinor index. We  identify $\lambda$ with the usual gluino of the MSSM and $\calo$ with  the fermion component of a color adjoint chiral super field containing its Dirac partner, the octino. Under  $U(1)_R$ symmetry, $\lambda$ and $\calo$  have opposite charges, $R=+1$ and $-1$, respectively.    If the $U(1)_R$ is exactly conserved by the mass terms, $\lambda$ and $\calo$ pair up to form a Dirac fermion,
\begin{equation}
\begin{aligned} 
-{\cal L}_{\rm mass}&=\frac12
\left(\lambda^\alpha~\calo^\alpha\right)
\left( \begin{array}{cc}
				0 		&  m_D \\
				m_D	& 0
		\end{array}\right)\left( \begin{array}{c}
												\lambda_\alpha \\ 
												\calo_\alpha
											\end{array}\right)+{\rm h.c.} \\
&=\frac12 m_D\left(\lambda^\alpha \calo_\alpha+\calo^\alpha\lambda_\alpha\right)+{\rm h.c.} \label{eq:lagdmass} \\
&=m_D\lambda \calo+m_D^\ast\lambda^\dagger O^\dagger.
\end{aligned}
\end{equation}
We are free to rotate $\lambda$ and $\calo$ such that the Dirac mass $m_D$ is real, and we do so. The Weyl spinors can be expressed in terms of creation and annihilation operators,
\begin{subequations}\label{eq:ints}
\begin{align}
\lambda_\alpha\left(x\right)&=\sum_s\int\frac{d^3{\bm p}}{\left(2\pi\right)^{3/2}\sqrt{2E_{\bm p}}}\left[x_\alpha\left({\bm p},s\right)a_{\bm p}^s e^{-ip\cdot x}\right.
\nonumber
\\
&\quad\quad\quad\quad\quad\quad\quad\quad\left.+y_\alpha\left({\bm p},s\right)b_{\bm p}^{s\dagger} e^{ip\cdot x}\right],
\\
\calo_\alpha\left(x\right)&=\sum_s\int\frac{d^3{\bm p}}{\left(2\pi\right)^{3/2}\sqrt{2E_{\bm p}}}\left[x_\alpha\left({\bm p},s\right)b_{\bm p}^s e^{-ip\cdot x}\right.
\nonumber
\\
&\quad\quad\quad\quad\quad\quad\quad\quad\left.+y_\alpha\left({\bm p},s\right)a_{\bm p}^{s\dagger} e^{ip\cdot x}\right],
\end{align}
\end{subequations}
where $E_{\bm p}=\sqrt{{\bm p}^2+m_D^2}$. $x$ and $y$ are momentum space solutions of the Dirac equation. Particle and antiparticle states are created by $a_{\bm p}^{s\dagger}$ and $b_{\bm p}^{s\dagger}$ respectively,
\begin{equation}
\begin{aligned} \label{eq:states}
\ket{{\bm p},s;\psi}&\equiv \left(2\pi\right)^{3/2}a_{\bm p}^{s\dagger}\ket{0},\\\ket{{\bm p},s;\bar\psi}&\equiv \left(2\pi\right)^{3/2}b_{\bm p}^{s\dagger}\ket{0}.
\end{aligned}
\end{equation}
Suppressing the spin index, we use $\ket{\psi}$ and $\ket{\bar\psi}$ to label the states as ${\bm p}\to0$. $\ket{\psi}$ carries $U(1)_R$ charge $R=-1$ while $\ket{\bar\psi}$ has $R=+1$.

\subsection{Hamiltonian}
In the nonrelativistic limit, the Hamiltonian in the $\left(\psi,\bar\psi\right)$ basis can be written as
\begin{align}
H_{ij}^{s^\prime,s}\equiv\bra{{\bm p}\to0,s^\prime;i}-{\cal L}_{\rm mass}\ket{{\bm p}\to0,s;j},
\end{align}
where $i,j=\psi,\bar\psi$ and $\mathcal{L}_{\rm mass}$ is given in Eq.~(\ref{eq:lagdmass}). Using the integral representations for $\lambda$ and $\calo$ from Eq.~(\ref{eq:ints}) the Hamiltonian becomes
\begin{align}
{\bm H}_D&=\left( \begin{array}{cc}
									m_D	&  0 						\\
									0		&  m_D
							\end{array} \right),
\end{align}
using $y^\alpha({\bm p},s')x_\alpha({\bm p},s)=m_D \delta_{s,s'}$ and suppressing the trivial dependence on spin and color. We have used the subscript $D$ here to emphasize that this is the Hamiltonian in the Dirac (conserved $U(1)_R$) case.

Next, we allow for the $U(1)_R$ symmetry to be slightly broken. This allows for small Majorana mass terms in the Lagrangian
\begin{equation}
\begin{aligned}
-\delta{\cal L}_{\rm mass}&=\frac12\left(m_\lambda\lambda\lambda+m_\calo \calo \calo\right) + {\rm h.c.},
\end{aligned}
\end{equation}
where we have suppressed the spinor indices. The Hamiltonian resulting from the Majorana mass terms can be found in the same way as in the Dirac case. The full Hamiltonian in the nonrelativistic limit, corresponding to ${\cal L}_{\rm mass}+\delta{\cal L}_{\rm mass}$, is
\begin{align}\label{eq:hmass}
{\bm H}&= {\bm H}_D+\delta {\bm H}=\left( \begin{array}{cc}
m_D &  m_M \\
m_M^\ast &  m_D
\end{array}\right),
\end{align}
where we have defined $m_M\equiv \left(m_\lambda^\ast+m_\calo\right)/2$. The eigenvalues of this Hamiltonian are
\begin{align}
M_{1,2}=m_D\pm\left|m_M\right|,
\end{align}
corresponding to the eigenstates
\begin{align}
\frac{\ket \psi\pm e^{-i\phi}\ket{\bar\psi}}{\sqrt2},
\end{align}
with $\phi=\arg\left(m_M\right)$.

\subsection{Interactions}
Now we would like to examine what happens to the Hamiltonian when we allow for  interactions of the Weyl fermions, in particular if they are allowed to decay. As a simple example, we consider a toy model which captures the essential physics. For now, we   consider the case where $\lambda$ and $\calo$ have Yukawa couplings to a fermion $\bar{d}$ and a complex scalar $\phi$ which are both fundamentals under color $SU(3)$,
\begin{align}
\mathcal{L}_{\rm int}=-\phi^\ast\left( y_\lambda\lambda^a+y_\calo \calo^a\right)t^a\bar{d}+{\rm h.c.},
\label{eq:Yuk}
\end{align}
where $t^a$ is a generator in the fundamental of $SU(3)$ normalized so that ${\rm tr}\left(t^at^b\right)=\delta^{ab}/2$ and $a$ labels the color of the adjoints. We will take $\bar{d}$ to be massless. If both $y_\lambda$ and $y_\calo$ are nonzero, $\mathcal{L}_{\rm int}$ breaks the $U(1)_R$.

\begin{figure}
\includegraphics[width=\linewidth]{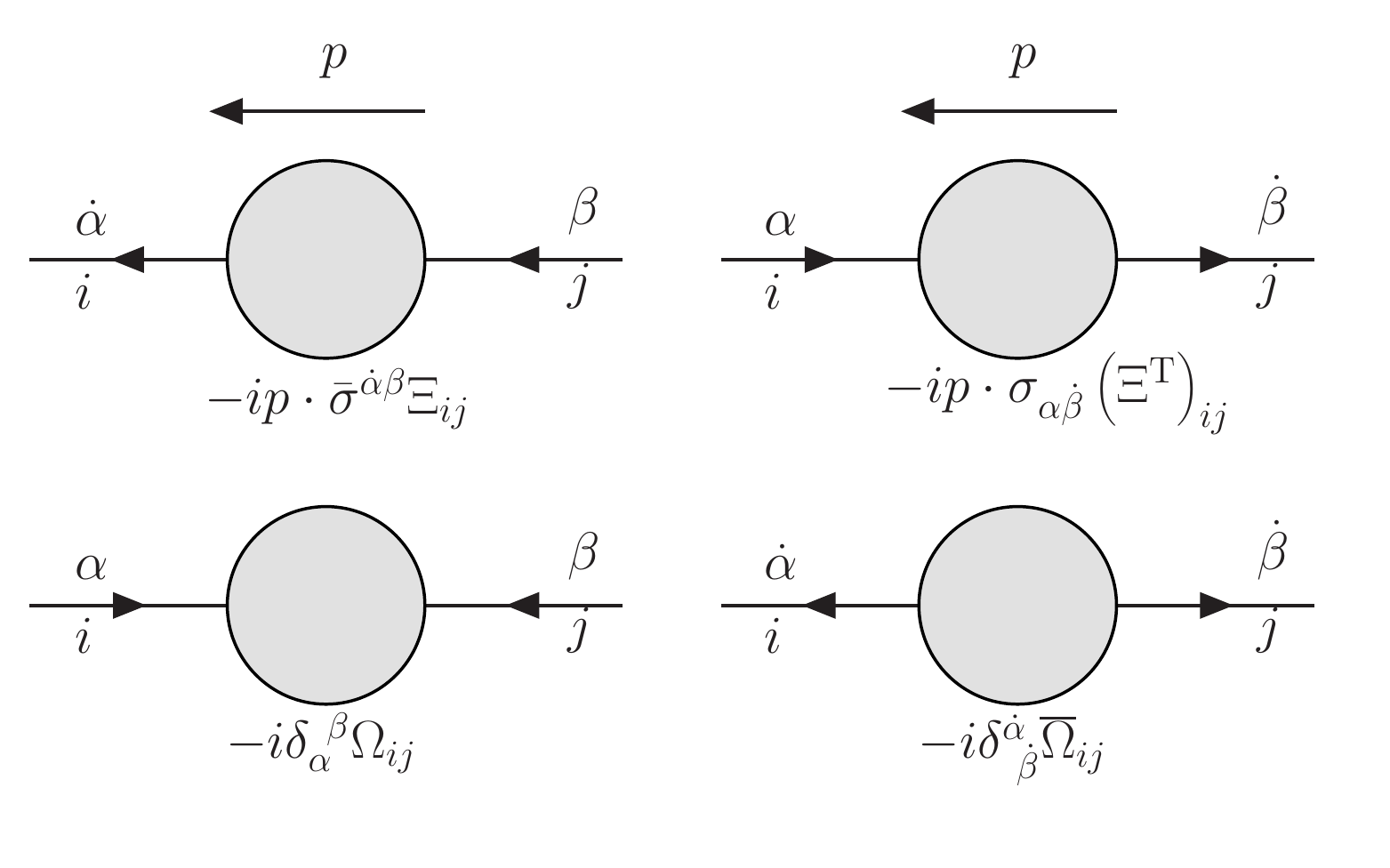}
\caption{Definitions of the self-energy functions for $i,j=\lambda,\calo$. The shaded circles represent the sum of all one-particle  irreducible, connected Feynman diagrams. External legs are amputated. $\alpha$ and $\beta$ are spinor indices. Arrows (and dots) denote left- or right-handed chiralities.}\label{fig:2ptfunc}
\end{figure}

With these interactions, the tree-level masses of $\ket{\psi}$ and $\ket{\bar\psi}$ ($M_{1,2}$) are modified (and possibly complex). As shown in~\cite{Dreiner:2010}, they are given by values of $\sqrt s$ that satisfy
\begin{equation}\label{eq:pmass}
\begin{aligned}
&\det\left[s{\bm 1}-\left({\bm 1}-{\bm\Xi}^{\rm T}\right)^{-1}\left({\bm m}+{\bm\Omega}\right)\right.
\\
&\quad\quad\quad\quad\quad\quad\quad\times\left.\left({\bm 1}-{\bm\Xi}\right)^{-1}\left(\overline{\bm m}+\overline{\bm\Omega}\right)\right]=0.
\end{aligned}
\end{equation}
In the expression above $\bm m$ is the tree-level fermion mass matrix in the $\lambda,\calo$ basis,
\begin{align}
{\bm m}= \left(\begin{array}{cc}
    m_\lambda & m_D \\ 
    m_D & m_\calo \\ 
  \end{array}\right),
\end{align}
and $\overline{\bm m}={\bm m}^\ast$.
${\bm \Xi}$ and ${\bm \Omega}$ are chirality-preserving and {-flipping} self-energy functions, respectively. They are shown in Fig.~\ref{fig:2ptfunc} along with the related functions ${\bm \Xi}^{\rm T}$ and $\overline{\bm \Omega}$. Note that these represent the finite pieces of the two-point functions (in some renormalization scheme); infinities in ${\bm \Omega}$ are absorbed by mass counterterms while those in ${\bm \Xi}$ are removed by wavefunction renormalization.

Corrections to the mass matrices are fixed by Eq.~(\ref{eq:pmass}) at leading order to be
\begin{equation}
\label{eq:masscor}
\begin{aligned}
  {\bm m}\big|_{\rm 1-loop} &= {\bm m} + {\bm \Omega} + \frac12\left({\bm m}\,{\bm \Xi} + {\bm \Xi}^{\rm T}{\bm m}\right), \\
  \overline{\bm m}\big|_{\rm 1-loop} &= \overline{\bm m} + \overline{\bm \Omega} + \frac12\left(\overline{\bm m}\,{\bm \Xi}^{\rm T} + {\bm \Xi}\,\overline{\bm m}\right).
\end{aligned}
\end{equation}
Armed with these expressions for the corrections to the mass matrices, we are ready to find the Hamiltonian for our toy model at one loop.

\begin{figure}
\includegraphics[width=\linewidth]{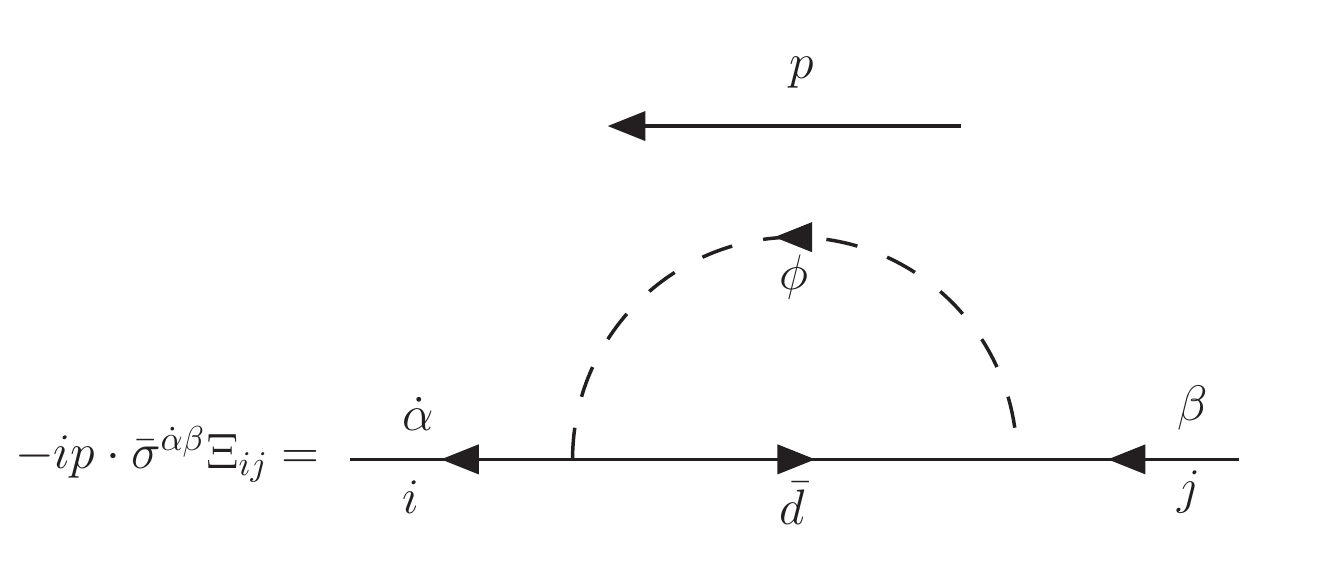}
\caption{The one-loop contribution to $\bm\Xi$ for $i,j=\lambda,\calo$ arising from the Yukawa interaction in Eq.~(\ref{eq:Yuk}). Contributions to the chirality-flipping two-point function $\bm\Omega$ are proportional to $m_{\bar d}$ which we neglect.}\label{fig:oneloop}
\end{figure}

In the toy model, ${\bm\Omega}\propto m_{\bar d}$ which we take to be vanishing so we are free to ignore ${\bm\Omega}$ and $\overline{\bm \Omega}$. In the $\overline{\rm MS}$ scheme, the elements of ${\bm\Xi}$, given by the diagram shown in Fig.~\ref{fig:oneloop}, are
\begin{equation}
\begin{aligned}
\Xi_{ij}&=\frac{y_i y_j^\ast}{4\left(4\pi\right)^2}\left[\left(1-\frac{m_\phi^2}{p^2}\right)\int_0^1dx\log\frac{\Delta}{Q^2}\right.
\\
&\quad\quad\quad\quad\quad\quad\quad\left.-\frac{m_\phi^2}{p^2}\left(1-\log\frac{m_\phi^2}{Q^2}\right)\right]
\end{aligned}
\end{equation}
with
\begin{align}
\Delta&=xm_\phi^2-x\left(1-x\right)p^2,
\end{align}
where $p$ is the momentum flowing through the diagram, $Q^2$ is the renormalization scale, and $i,j=\lambda,\calo$.

If $m_\phi^2<p^2$, there are on-shell intermediate states that give rise to an imaginary part in the loop integral for $\Xi_{ij}$,
\begin{align}
\Im\left(\frac{\Xi_{ij}}{y_iy_j^*}\right)=\frac{1}{(4\pi)^2}\frac{\pi}{4}\left(1-\frac{m_\phi^2}{p^2}\right)^2\theta\left(p^2-m_\phi^2\right).
\end{align}
$(\Xi^{\rm T})_{ij}$ is obtained from $\Xi_{ij}$ through the relation $(\Xi^{\rm T})_{ij}=\Xi_{ij}^\star$, where $\star$ means taking the complex conjugate of the Lagrangian parameters but not of integrals over loop momenta. We express the one-loop mass matrices in Eq.~(\ref{eq:masscor}) in terms of the elements of ${\bm\Xi}$ and ${\bm \Xi}^{\rm T}$ which we use to find the Hamiltonian at one loop and to leading order in $m_{\lambda,\calo}/m_D$ using Eq.~(\ref{eq:hmass}),
\begin{align}
{\bm H}&=
\left( 
\begin{array}{cc}
m_D+\delta_D & m_M+\delta_M  \\
m_M^\ast+\delta_M^\ast &  m_D+\delta_D
\end{array}
\right), \label{Eq:fulltoyH}
\end{align}
with 
\begin{equation}
\begin{aligned}
 \delta_D&=\frac{m_D}{4}\left(\Xi_{\lambda\lambda}+\Xi_{\calo\calo}+\Xi_{\lambda\lambda}^{\rm T}+\Xi_{\calo\calo}^{\rm T}\right),
 \\ 
 \delta_M&=\frac{m_D}{2}\left(\Xi_{\calo\lambda}+\Xi_{\lambda\calo}^{\rm T}\right),
 \\ 
 \delta_M^\ast&=\frac{m_D}{2}\left(\Xi_{\lambda\calo}+\Xi_{\calo\lambda}^{\rm T}\right),
\end{aligned}
\label{eq:corrections}
\end{equation}
where we evaluate the one-loop diagrams at the scale of the fermion masses, $p^2\simeq m_D^2$. The dispersive parts of $\delta_D$ and $\delta_M^{(\ast)}$ are corrections to the Dirac and Majorana masses, respectively, while the absorptive parts arising from on-shell intermediate states are related to the decays of the pseudo-Dirac fermions. As in the purely Dirac case, $m_D$ is multiplicatively renormalized while the Majorana masses pick up corrections proportional to the Dirac mass times a $U(1)_R$-breaking combination of couplings. 

We can separate the Hamiltonian into its dispersive and absorptive parts in the standard way,\footnote{The form of the Hamiltonian we will arrive at differs from the Hamiltonian in Eq.~(7) of Ref.~\cite{Grossman:2012nn}.}
\begin{align}
{\bm H}&={\bm M}-\frac{i}{2}{\bm\Gamma},
\label{eq:fullH}
\end{align}
where we set
\begin{align}
{\bm M}&=
\left( 
\begin{array}{cc}
M_D &  M_M \\
M_M^\ast &  M_D
\end{array}
\right).
\end{align}
$M_D$ and $M_M$ are given by $m_D$ and $m_M$ plus the dispersive parts of the one-loop corrections in Eq.~(\ref{eq:corrections}), with the renormalization condition that the pseudo-Dirac fermions have pole masses $M_D\pm\left|M_M\right|$ (in the limit that the width difference can be ignored). Since we have rotated $\lambda$ and $\calo$ so that $m_D$ is real, $M_D$ is real. From the structure of the one-loop corrections in Eq.~(\ref{eq:corrections}), $M_D\simeq m_D$ and we expect that in the absence of fine-tuning,
\begin{align}
\left|M_M\right|\gtrsim\frac{\left|y_\lambda y_\calo^\ast\right|}{\left(4\pi\right)^2}M_D.
\end{align}
The absorptive part of the Hamiltonian is
\begin{equation}
\begin{aligned}
{\bm\Gamma}&\simeq\frac{M_D}{64\pi}\left(1-\frac{m_\phi^2}{M_D^2}\right)^2
\\
&\quad\times\left( 
\begin{array}{cc}
\left|y_\lambda\right|^2+\left|y_\calo\right|^2 &  2 y_\lambda y_\calo^\ast \\
2 y_\lambda^\ast y_\calo &  \left|y_\lambda\right|^2+\left|y_\calo\right|^2
\end{array}
\right).
\end{aligned}
\end{equation}
As written, there are three phases in $\bm H$ but only one combination is physical since we have the freedom to remove two. For example, we can rotate one linear combination of $\lambda$ and $\calo$ so that $M_M$ is real (another linear combination was rotated to make $M_D$ real) and by rotating $\phi^\dagger \bar d$ we can make $y_\lambda$ or $y_\calo$ real but not necessarily both simultaneously.

\subsection{Oscillations}
The form of $\bm H$ in Eq.~(\ref{eq:fullH}) is the same as the two-state Hamiltonians relevant to neutral meson mixing. Therefore, we can simply adapt the same formalism to study the oscillations of the pseudo-Dirac fermions. We briefly review some of this formalism from Ref.~\cite{Beringer:1900zz}. For a more general treatment of oscillations see~\cite{Martone:2011kh,*Pilaftsis:1997dr}.

In terms of the states $\ket{\psi}$ and $\ket{\bar{\psi}}$ defined in Eq.~(\ref{eq:states}), the eigenstates of $\bm H$ are 
\begin{align}
\ket{\psi_H} &=p\ket{\psi}-q\ket{\bar{\psi}},\quad
\ket{\psi_L} =p\ket{\psi}+q\ket{\bar{\psi}},
\end{align}
with eigenvalues $\omega_{H,L}$. The subscripts $H$ and $L$ refer to the heavy and light mass states respectively with masses $m_{H,L}$, and 
\begin{align}
\left(\frac{q}{p}\right)^2=\frac{M_{12}^*-(i/2)\Gamma_{12}^*}{M_{12}-(i/2)\Gamma_{12}}.
\end{align}
where $M_{12}$ and $\Gamma_{12}$ are the 1-2 elements of $\bm M$ and $\bm\Gamma$. \emph{CP} violation in mixing occurs when $\left|q/p\right|\ne 1$. The mass and width differences $\Delta m$ and $\Delta\Gamma$ between the two eigenstates are 
\begin{equation}
\begin{aligned}
\Delta m&=m_H-m_L=\Re(\omega_H-\omega_L),\\
\Delta\Gamma&=\Gamma_H-\Gamma_L=-2\Im(\omega_H-\omega_L),
\end{aligned}
\end{equation}
where
\begin{align}
\omega_H-\omega_L=2\sqrt{\left(M_{12}-\frac{i}{2}\Gamma_{12}\right)\left(M_{12}^*-\frac{i}{2}\Gamma_{12}^*\right)}.
\end{align}

A state that is initially pure $\ket{\psi}$ or $\ket{\bar{\psi}}$ evolves in time to a mixture of both states due to oscillations as follows\footnote{Note that in Ref.~\cite{Grossman:2012nn}  the case of neutralino oscillations with   fairly long lifetimes    was discussed, and it was claimed that an oscillating decay rate could be an interesting consequence. Our analysis shows that in the absence of \emph{CP} violation  the decay rate does not oscillate in the rest frame, and therefore, by Lorentz invariance, will not oscillate for boosted particles either. Instead it is possible for the particle content of the final states to oscillate.}
\begin{equation}\label{eq:evolve}
\begin{aligned}
\ket{\psi(t)} &= g_+(t)\ket{\psi}-\frac{q}{p}\,g_-(t)\ket{\bar{\psi}}, \\
\ket{\bar{\psi}(t)}&=g_+(t)\ket{\bar{\psi}}-\frac{p}{q}\,g_-(t)\ket{\psi},
\end{aligned}
\end{equation}
where
\begin{align}
g_\pm(t)=\frac12\left(e^{-im_Ht-\frac12\Gamma_H t}\pm e^{-im_Lt-\frac12\Gamma_L t}\right).
\end{align}

To characterize the oscillations, it is often useful to define two dimensionless parameters,
\begin{align}\label{eq:x}
x\equiv\frac{\Delta m}{\Gamma},~y\equiv\frac{\Delta \Gamma}{2\Gamma}.
\end{align}
If $x\ll1$, the states decay before oscillating while if $x\gg1$, the states rapidly oscillate before decaying, making it difficult to observe oscillation signatures. The effects of oscillations are maximized for $x\sim1$. For $\left|y\right|$ near unity (as defined $\left|y\right|\le 1$) one of the two states can be rapidly depleted before the other decays, as is the case in the kaon system. If $\left|y\right|\ll1$, neither state is preferentially depleted over the other.  

\section{A Specific Example} \label{sec:model}
\subsection{UV Theory}
We work with a nearly $U(1)_R$ symmetric SUSY model. The left-handed gauginos and the scalar superpartners of left-handed fermions have $R$ charge +1, while the SM particles have $R$ charge 0.   We take the SM left-handed Weyl fermions $q_i,\bar{u}_i,\bar{d}_i,\ell_i,\bar{e}_i$ to be  components of left chiral superfields $\Phi_{q_i}, \Phi_{\bar{u}_i}, \Phi_{\bar{d}_i}, \Phi_{\ell_i}, \Phi_{\bar{e}_i}$. The gluino $\lambda$ is the fermion  component of the QCD field strength superfield $ W_{c}^{\alpha}$. We assume the gluino to be the lightest superpartner other than the gravitino. In order for the gluino to get a $U(1)_R$ preserving Dirac mass, we introduce a left chiral, color adjoint superfield
$\Phi_{\calo}$ whose fermion  component $\calo$ is the Dirac partner of the gluino. For simplicity, we do not discuss the Higgs or the electroweak sectors in this work, but we note that it is possible to build a viable model with an extended Higgs sector and/or  lepton number violation which preserves the $U(1)_R$ symmetry~\cite{Hall:1990hq,*Kribs:2007ac,Frugiuele:2012kp,*Bertuzzo:2014bwa,*Davies:2011mp,*Frugiuele:2011mh,*Frugiuele:2012pe,*Riva:2012hz}. In order to allow   non gauge  interactions for $\Phi_\calo$, we introduce superfields   $\Phi_{\bar{D} }$ and $\Phi_D$, transforming under the SM gauge group in the same way as $\bar{d}$ and $\bar{d}^*$  respectively. We show the field content of the model that is most relevant to our study in Table~\ref{table:particles}.

These fields have a  superpotential mass term and  interactions
\begin{equation}\label{eq:dsuper}
\int d^2\theta\  \mu_D  \Phi_{\bar{D}}\Phi_D + y \Phi_{\bar{D}}\Phi_\calo\Phi_D+ g^\prime_i\Phi_{\bar{d}_i}\Phi_\calo\Phi_D + {\rm h.c.},\end{equation} where $ \mu_D$ is assumed to be very large, of order a TeV or higher.  We neglect the possibility of mixing between the ordinary down quarks and the fermion components of $\Phi_{\bar{D}},\Phi_D$.

\begin{table}
    \begin{tabular}{ | c | c | c |r| c | }
    \hline
     Field & $SU(3)$ & $SU(2)$ & $U(1)$ & $U(1)_R$ \\ \hline \hline
     $q_i$ & 3 & 2 & $1/6$ & 0 \\  
     $\bar{u}_i$ & $\bar{3}$ & 1 & $-2/3$ & 0 \\ 
     $\bar{d}_i$ & $\bar{3}$ & 1 & $1/3$ & 0 \\  
     $\ell_i$ & 1 & 2 & $-1/2$ & 0 \\ 
     $\bar{e}_i$ & 1 & 1 & 1 &  0 \\  
     $\lambda$	& 8 & 1 & 0 & $+1$ \\ 
     $\calo$ & 8 & 1 & 0 & $-1$ \\ 
     $\phi_{\bar{d}_i}$ & $\bar{3}$	& 1 & $1/3$	& $+1$ \\ 
     $\phi_{\bar{D}}$ & $\bar{3}$ & 1 & $1/3$ & $+1$ \\ 
	$\phi_{D}$ & ${3}$ & 1 & $-1/3$	& $+1$ \\ \hline
    \end{tabular} 
    \caption{Part of the particle content of the model with quantum numbers under the SM gauge group and $U(1)_R$. All fermions are left-handed Weyl spinors. $\lambda$ is the gluino, and $\calo$ is the octino. The $\phi_{\bar{d}}$ fields are  scalar superpartners of SM quarks, and  $\phi_D,\phi_{\bar{D}}$ are superpartners of exotic heavy vector-like quarks. The fields $q_i,\bar{u}_i,\bar{d}_i,\ell_i,\bar{e}_i$ are SM  fermions and $i$ is a generational index.} \label{table:particles}
\end{table}

 We assume that the gluino is the lightest $R$-charged particle and   decays via $U(1)_R$ symmetry violation. $U(1)_R$ symmetry must be broken by supergravity, and we will assume that R-parity is also broken. There is an extensive literature on R-parity violating interactions  and their phenomenological constraints \cite{Ellis:1984gi,*Ross:1984yg,Barbier:2004ez}. In this example, to ensure proton stability, we will assume that baryon number is conserved. We include the following  R-parity  and $U(1)_R$-symmetry--violating superpotential terms:
\begin{equation}
\begin{aligned}
\int d^2\theta \ y_{ijk} &\Phi_{\ell_i}\Phi_{q_j}\Phi_{\bar{d}_k}+y'_{ij} \Phi_{\ell_i }\Phi_{q_j}\Phi_{\bar{D}}
\\
&+ y''_{ij}\Phi_{\bar{e}_i}\Phi_{\bar{u}_j}\Phi_D +{\rm h.c.}
\end{aligned}
\end{equation}
The first two terms leave  the linear combination $R-L$ unbroken, where $L$ is lepton number, and  the third term leaves $R+L$ unbroken.

We do not discuss a specific SUSY-breaking model here, but  assume that SUSY is broken in a hidden sector which communicates with the visible sector at the messenger scale $\Lambda_M$.   Supersymmetry breaking is incorporated via spurions $W'_\alpha$ and $X$, where $W'_\alpha$ is the expectation value of a hidden sector $U(1)$ gauge field strength,  and $X$ is the expectation value of a hidden sector  chiral superfield.  We set
\begin{equation}
\begin{aligned}
W'_\alpha=&D\theta_\alpha ,
\\ 
X=& F\theta^2 ,
\end{aligned}
\end{equation}
where $D$ and $F$ are SUSY-breaking order parameters which are $U(1)_R$ neutral. We assume that   $X$ transforms nontrivially under some symmetry of the SUSY-breaking sector. Because $X$ is not a singlet, there can be no $U(1)_R$-symmetry--violating Majorana gaugino mass terms from spurions  such as 
$\int d^2\theta \left(X/\Lambda_M\right) W_\alpha W^\alpha$ where $W_\alpha$ is a SM gauge field strength superfield.   The Dirac gluino mass arises from the spurion term
\begin{equation}\int d^2\theta \frac{c\ W'_\alpha }{ \Lambda_M} W_{c}^\alpha \Phi_\calo  + {\rm h.c.},\end{equation}  where $c$ is a dimensionless parameter, giving 
\begin{equation}
\label{eq:diracmass}
m_D=\frac{c  D}{\Lambda_M} \ .
\end{equation}
Majorana mass terms for the gauginos   and scalar $\phi_D$, $\phi_{\bar{D}}$ mixing will  be generated from anomaly mediation~\cite{Randall:1998uk,*Giudice:1998xp}, which gives, e.g.  a Majorana gluino mass, 
\begin{equation}
\label{eq:mlambda}
m_\lambda=\frac{\beta_s}{g_s} m_{3/2} \ ,\end{equation}  
and scalar mass mixing term,
\begin{equation}\label{bterm}
m_{3/2}  \mu_D\phi_D \phi_{\bar{D}}+ {\rm h.c.},\end{equation}
 where $\beta_s$ is the beta function for the QCD coupling $g_s$. $m_{3/2}\sim\left(D+F\right)/M_{\rm Pl}$ is the gravitino mass with $M_{\rm Pl}$ the Planck scale. Note that we must assume that $m_{3/2}$ is small in order to have an approximate $U(1)_R$ symmetry, so   $\Lambda_M$ must be well below the Planck scale. 
  The spurion $X$ can give rise to SUSY-breaking scalar masses via operators  
\begin{equation}
\begin{aligned}
\int d^4\theta \frac{X^\dagger X}{\Lambda_M^2} &\left(c_{ij}\Phi^\dagger_{\bar{d}_i}\Phi_{\bar{d}_j}+ c_{i}\Phi^\dagger_{\bar{D}}\Phi_{\bar{d}_i}\right.\\ &\left.+c_{\bar{D}}\Phi^\dagger_{\bar{D}}\Phi_{\bar{D}}+ c_D\Phi^\dagger_D\Phi_D\right) \ ,
\end{aligned}
\end{equation}
where $c_{ij},c_i,c_D,c_{\bar{D}}$ are dimensionless parameters. We will assume a modest hierarchy of supersymmetry breaking terms,
\begin{equation}
D< F,
\end{equation}
so that in general scalar masses are larger than gaugino masses. 
A Majorana mass term for $\calo$  could arise  from a  $U(1)_R$-violating superpotential term
\begin{equation}
\label{eq:mo}
\int d^2\theta\  m_\calo  \Phi_\calo^2+ {\rm h.c.}
\end{equation} 
Note that  as we assume all $U(1)_R$-violating terms are small, \begin{equation} m_\calo\ll m_D . \end{equation} A possible explanation for the small size of this term is that   $\calo$  could be part of an approximately $\mathcal{N}=2$ supersymmetric gauge or gauge/Higgs sector \cite{Fayet:1975yi,Fayet:1984jt,Fox:2002bu}. 

Supersymmetry breaking may also provide the supersoft terms \cite{Fox:2002bu}
\begin{align}
\int d^2\theta \frac{{W'}^\alpha W'_\alpha}{\Lambda_M^2}&\big( c_{\calo\calo} \Phi_\calo^2\\
&+c_{D\bar{D}}  \Phi_D\Phi_{\bar{D}}+c_{D{\bar{d}}_i} \Phi_D\Phi_{{\bar{d}}_i}\big)+ {\rm h.c.},\nonumber
\end{align}
where  $c_{\calo\calo},c_{D\bar{D}},c_{D{\bar{d}}_i}$ are dimensionless parameters. These give scalar mass mixing terms, including
\begin{equation}\label{eq:btermtwo}
 \left(B^2_{D\bar{D} } \phi_D \phi_{\bar{D}}+   B^2_{D{\bar{d}}_i } \phi_D \phi_{{\bar{d}}_i}\right)+{\rm h.c.},\end{equation}
with 
\begin{equation}
 B^2_{D\bar{D} }=\frac{c_{D\bar{D}} D^2}{ \Lambda_M^2}, \ \ \   B^2_{D{\bar{d}}_i }=\frac{c_{D{\bar{d}}_i }D^2}{ \Lambda_M^2}\ .\end{equation}

Our assumption of relatively large $F$ term contributions to scalar masses solves the   negative scalar mass squared problem \cite{Csaki:2013fla,*Benakli:2008pg,*Benakli:2010gi}, preserves the $U(1)_R$ solution to the SUSY \emph{CP} problem, but does not address the SUSY flavor problem.  We assume the latter is addressed by an  approximate flavor symmetry of the messenger interactions, leading to
\begin{equation}
 c_{ij}=c_{\bar{d}} \delta_{ij} + {\rm small} \end{equation}  and 
 \begin{equation}
 c_i,c_{D{\bar{d}}_i}\ll  c_{\bar{d}} \ .  \end{equation} 
  Thus the  $\phi_{\bar{d}_k}$ are nearly degenerate, with mass squared
   \begin{equation}  m_{\bar{\phi}_{d_k}}^2\approx \frac{c_{\bar{d}}    F^2}{\Lambda_M^2}\ .\end{equation} We assume 
   \begin{equation} \mu_D > \frac{F}{\Lambda_M} \end{equation} so that $\phi_D,\phi_{\bar{D}}$ are nearly degenerate with mass $\mu_D$.
  We also note that there is a supersymmetry-breaking mixing term between the $\phi_{\bar{d}_k}$ and $\phi_{\bar{D}}$,
\begin{equation}
\call\supset -{\tilde{m}}_{k}^2 \phi_{\bar{d}_k} \phi_{\bar{D}}^*+{\rm h.c.},
\end{equation} with   
\begin{equation}
 {\tilde{m}}_{k}^2=\frac{c_k F^2}{\Lambda_M^2}
\end{equation} assumed to be small.

The supersymmetric gauge interactions contain the Yukawa couplings
\begin{equation}
\call\supset -\sqrt2 g_s\bar{d}_i\lambda^a t^a \phi_{\bar{d}_i}^* +{\rm h.c.}
\end{equation} 
 and the superpotential terms contain  the interactions
\begin{equation}
\begin{aligned}
\call&\supset -g'_i\bar{d}_i\calo^a t^a\phi_D-y_{ijk} \ell_iq_j\phi_{\bar{d}_k}
\\
&\quad\quad-y^{\prime}_{ij} \ell_i q_j\phi_{\bar{D}} - y^{\prime\prime}_{ij}\bar{e}_i^\ast\bar{u}_j^\ast\phi_D^\ast+{\rm h.c.}
\end{aligned}
\end{equation}

While the flavor-violating terms $B^2_{D{\bar{d}}_k}$ and $\tilde m_k^2$ can be suppressed at tree-level by a flavor symmetry as we assume, they are generated at one-loop and proportional to $y_{ijk}y^{\prime}_{ij}B^2_{D\bar{D}}$ and $y_{ijk}y^{\prime}_{ij}$, respectively.

 \subsection{Effective Four Fermion Theory for Gluino Decays}

Now we assume that all the  squarks are heavy and can be integrated out. Using a mass insertion approximation for the small scalar mixing terms and neglecting the gravitino mass,  the resulting effective  4-fermi Lagrangian for the gluino interactions is approximately
\begin{equation}\label{eq:Leff}
\begin{aligned}
\call_{\rm eff}&=G_{ijk} \lambda \ell_i q_j \bar{d}_k + G^\prime_{ijk} \calo  \ell_i q_j\bar{d}_k + 
\\
&\quad\quad G^{\prime\prime} _{ijk}\calo \bar{e}^*_i\bar{u}^*_j\bar d_k+G^{\prime\prime\prime}_{ijk}\lambda \bar{e}^*_i\bar{u}^*_j\bar d_k +{\rm h.c.},
\end{aligned}
\end{equation}
where we have suppressed color indices and
\begin{align}
G_{ijk}&=\frac{\sqrt2g_s y_{ijk}}{m_{\phi_{\bar d}}^2}\ ,\ \ \ G^\prime_{ijk}=\frac{g_k^\prime y_{ij}^\prime B^2_{D\bar{D} }}{\mu_D^4} \ ,\\ 
G^{\prime\prime}_{ijk}&=\frac{g_k^\prime y_{ij}^{\prime\prime}}{ \mu_D^2}\ ,\ \ G^{\prime\prime\prime}_{ijk}=\frac{\sqrt2g_s y^{\prime\prime}_{ij} ({\tilde{m}}_{k}^2  B^2_{D\bar{D} }+\mu_D^2B^2_{D{\bar{d}}_k })}{m_{\phi_{\bar d}}^2\mu_D^4} \ .\nonumber
\end{align}
We have assumed a flavor symmetry such that $B^2_{D{\bar{d}}_k}$ and $\tilde{m}_{k}^2$ are loop-suppressed so we might expect that $G^{\prime\prime\prime}_{ijk}$ is somewhat smaller than the others.

\subsection{Pseudo-Dirac Gluino Oscillations} \label{sec:pdgluino}
\begin{figure}
\includegraphics[width=\linewidth]{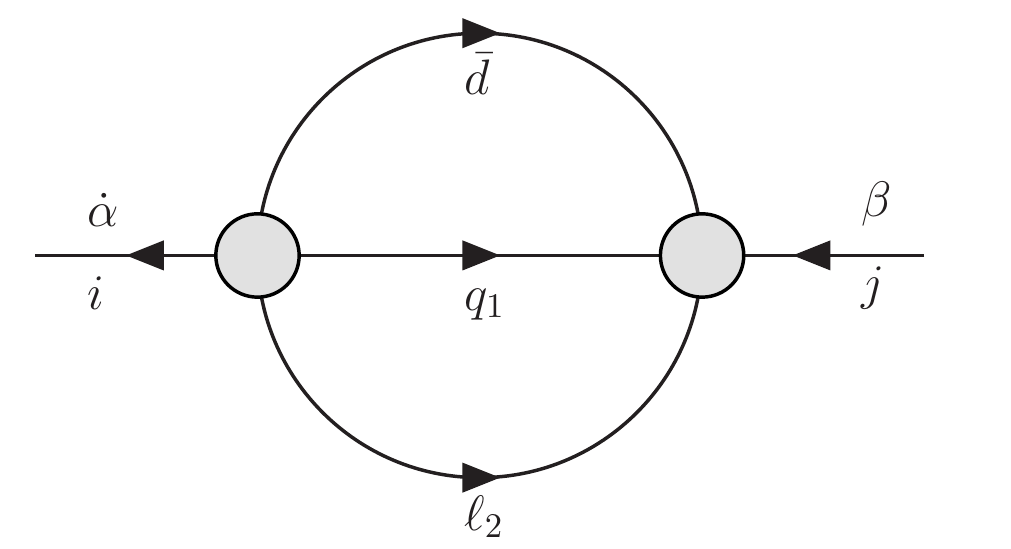} 
\caption{Two-loop corrections to the two-point functions, $\Xi_{ij}$, for $i,j=\lambda, \calo$, that arise due to the couplings in Eq.~(\ref{eq:Leff2}).}\label{fig:ff2loop}
\end{figure}

In this section we use the machinery from Sec.~\ref{sec:pdfermion} with our specific model. In the toy model the scalar field $\phi$ was light so that the gluino decayed to a scalar and a fermion. However, in this specific model the squarks are heavier than the gluino so the relevant decays are to 3-body final states through the 4-fermi operators in Eq.~(\ref{eq:Leff}). Therefore, the one-loop corrections to the gluino self-energy (as seen in Fig.~\ref{fig:oneloop}) are real. Absorptive contributions to the Hamiltonian occur at two-loop order through  diagrams like the one shown in Fig.~\ref{fig:ff2loop}. We continue to ignore the chirality-flipping two-point functions, $\mathbf{\Omega}$, since they are proportional to light fermion masses.

For simplicity we will consider an example where there are only two relevant 4-fermi operators, assuming that $G_{211}\equiv\tilde{G}_\lambda$ and $G^{\prime}_{211}\equiv\tilde{G}_\calo$ dominate in Eq.~(\ref{eq:Leff}),\footnote{Due to our assumption of a flavor symmetry, we expect that $G^{\prime\prime\prime}_{ijk}$ is loop-suppressed. Since \emph{CP}-violating effects involving the lepton singlet final state are proportional to $G^{\prime\prime}G^{\prime\prime\prime}$, this makes the lepton doublet final state that we have chosen more interesting.} leading to the effective Lagrangian
\begin{equation}\label{eq:Leff2}
\call_{\rm eff}= \tilde{G}_\lambda \lambda \bar{d}q_1\ell_2+ \tilde{G}_\calo \calo \bar{d}q_1\ell_2+{\rm h.c.},
\end{equation}
suppressing gauge indices.

The imaginary part of the diagram in Fig.~\ref{fig:ff2loop} is found to be
\begin{align}
\Im\left(\frac{\Xi_{ij}}{\tilde{G}_i\tilde{G}_j^\ast}\right)=\frac{2p^4}{3\left(16\pi\right)^3}
\end{align}  
for $i,j=\lambda,\calo$. Following the discussion in Sec.~\ref{sec:pdfermion}, in the presence of these interactions, the Hamiltonian for the pseudo-Dirac gluino is
\begin{align}
{\bm H}={\bm M}-\frac{i}{2}{\bm \Gamma}
\end{align}
with
\begin{equation}\label{eq:pdgham}
\begin{aligned}
&{\bm M}=
\left(\begin{array}{cc}
M_D & M_M \\
M_M^\ast & M_D
\end{array}\right),
\\ 
&{\bm \Gamma}\simeq\left(\begin{array}{cc}
\Gamma_0 & 0 \\
0 & \Gamma_0
\end{array}\right)
\\ 
&+\frac{M_D^5}{12\left(8\pi\right)^3} 
\left(\begin{array}{cc}
|\tilde{G}_\lambda|^2+|\tilde{G}_\calo|^2 & 2\tilde{G}_\lambda^\ast\tilde{G}_\calo \\
2\tilde{G}_\lambda\tilde{G}_\calo^\ast & |\tilde{G}_\lambda|^2+|\tilde{G}_\calo|^2
\end{array}\right).
\end{aligned}
\end{equation}
$\Gamma_0$ represents possible contributions to the decay width that involve $\lambda$ (or possibly $\calo$) that do not arise from operators in Eq.~(\ref{eq:Leff2}) and do not break the $U(1)_R$ symmetry, such as decays to a gluon and gravitino. The masses in $\bm M$ are the renormalized two-loop values such that the gluino and octino form nearly Dirac fermions with masses $M_D\pm\left|M_M\right|$ (ignoring any width difference). The Dirac mass is multiplicatively renormalized from its tree-level value in Eq.~(\ref{eq:diracmass}), leading to $M_D\simeq m_D$. At tree-level the Majorana mass, $M_M$, is $m_\lambda^\ast+m_\calo$ from Eqs.~(\ref{eq:mlambda}) and (\ref{eq:mo}). It also receives one-loop contributions proportional to the Dirac mass and $U(1)_R$-violating terms,
\begin{equation}
\delta_M\sim\frac{g_s g_k^\prime}{\left(4\pi\right)^2} \left(\frac{{\tilde{m}}_{k}^2  B^2_{D\bar{D} }+\mu_D^2B^2_{D{\bar{d}}_k }}{\mu_D^4}\right)M_D.
\end{equation}
Since we have assumed a flavor-symmetry so that $B^2_{D{\bar{d}}_k}$ and $\tilde{m}_{k}^2$ are loop-suppressed, this is effectively a two-loop contribution to the mass.

\section{\emph{CP} violation in pseudo-Dirac fermion  oscillations} \label{sec:CP}

Particle-antiparticle oscillations can enhance the observable effects of \emph{CP}-violating phases, via interference between the phases in oscillations and decay amplitudes. Here we consider a possible charge asymmetry between  like sign dimuons that may be produced in gluino decays as a possibly observable example. Like sign dileptons are a standard SUSY signal~\cite{Barnett:1993ea,*Guchait:1994zk,Baer:1991xs,*Baer:1995va} and can exhibit a charge asymmetry at a $pp$ collider like the LHC when the squarks are lighter than the gluino~\cite{Baer:1991xs,*Baer:1995va}, while in our scenario the gluino is lighter than the squarks. Since the interactions that produce a pair of pseudo-Dirac gluinos conserve $U(1)_R$,  initially a pair of $R=+1$ and $R=-1$ states are produced. We denote the amplitude for a state with $U(1)_R$ charge $R$ to decay to $\mu^\pm$ as $\calm_R^\pm$. Note that $\calm_+^+$ can arise from the couplings $G_{2jk} , G'''_{2jk} $ in Eq.~(\ref{eq:Leff}), while $\calm_-^-$ can arise from their hermitian conjugates. At tree level, and neglecting final state interactions, we expect no direct \emph{CP} violation, and may assume
\begin{equation}\label{eq:nodirect}
\calm_-^-={\calm_+^+}^\ast\ .\end{equation}
Similarly, $\calm_-^+$ is proportional to the couplings $G'_{2jk},G''_{2jk}$ and $\calm_+^-$ to their conjugates.  Assuming no direct \emph{CP} violation gives
\begin{equation}\label{eq:nodirectt}
\calm_-^+={\calm_+^-}^\ast\ .  \end{equation} 
\emph{CP} violation due to interference between a decay with mixing and without mixing is only possible when either $R$ charge state can decay into an indistinguishable final state.  Since the   $G,G'$ operators tend to produce different helicities than the $G'',G'''$ operators, \emph{CP} violation from interference will be maximized when either the $G$-$G'$ or the $G''$-$G'''$ pair dominate, and when the quark flavor dependence of  the different couplings is the same. In the following analysis we assume that the  final states produced by the $\calm_+^+$ decay amplitudes are indistinguishable from those produced by the $\calm_-^+$ amplitudes, and the final states from $\calm_+^-$ are indistinguishable from those due to $\calm_-^-$.

Using Eq.~(\ref{eq:evolve}), we find that a state with $R=+1$ at $t=0$ decays into $\mu^\pm$ at time $t$ with an amplitude of 
\begin{equation}
A_+^\pm(t) = g_+\left(t\right)\calm_+^\pm -\frac{p}{q}\,g_-\left(t\right)\calm_-^\pm \ ,
\end{equation}
and an initial $R=-1$ state decays into $\mu^\pm$ at time $t$ with an amplitude  of
\begin{equation}
A_-^\pm(t) = g_+\left(t\right)\calm_-^\pm -\frac{q}{p}\,g_-\left(t\right)\calm_+^\pm \ .
\end{equation}
It is possible that the oscillation length is too short to be directly observable for gluino decays. However, interference between decays with and without mixing can still produce sizable observable \emph{CP} violation when the oscillation and decay times are similar. Assuming initial incoherent production of a pair of gluinos with opposite $R$ charges,   the number of resulting like-sign pairs of  positively charged muons, $N^{++}$, versus negatively charged muons, $N^{--}$, where
\begin{align}\label{eq:mss}
N^{\pm\pm}\propto\left[\int_0^\infty dt\left|A_+^\pm(t)\right|^2\right]\times\left[\int_0^\infty dt\left|A_-^\pm(t)\right|^2\right],
\end{align}
can exhibit a nonzero asymmetry,
\begin{align}\label{eq:asymmetry}
A&\equiv\frac{N^{++}-N^{--}}{ N^{++}+ N^{--}}.
\end{align}
Also of interest is the total fraction of same sign muon decays,
\begin{align}\label{eq:fraction}
R&\equiv\frac{N^{++}+N^{--}}{ N^{+-}+ N^{-+}+N^{++}+ N^{--}},
\end{align}
where we calculate the number of opposite sign muon decays in an analogous way to Eq.~(\ref{eq:mss}). Below, we show approximate expressions for $A$ and $R$ in some physically relevant limits.

If the total decay width of the pseudo-Dirac particles is dominated by final states that do not include muons and do not break the $U(1)_R$, then we can ignore the width difference between the states, $\Delta\Gamma$, and take $|q/p|=1$. This corresponds to $\Gamma_0\gg\Gamma$ in Eq.~(\ref{eq:pdgham}). Then the asymmetry can be expressed as
\begin{align}
A&\simeq\frac{4 x   r (1-r^2)  \sin\beta}{ (1+x^2)^2(1+r^2)^2-(1-r^2)^2 -4x^2 r^2 \sin^2\beta},
\end{align}
where $x$ is related to the mass difference as in Eq.~(\ref{eq:x}) and
we assume no direct \emph{CP} violation as in Eqs.~(\ref{eq:nodirect}) and (\ref{eq:nodirectt}). We have defined a reparameterization-invariant phase,
 \begin{equation}
 \beta\equiv {\rm arg}\left( \frac{q}{p}  \frac{\calm_+^+}{\calm_-^+}  \right),
 \end{equation} 
and a ratio of amplitudes,
 \begin{equation}r\equiv \frac{|\calm_+^+|}{|\calm_-^+|}\ .\end{equation} 
In the same limit the ratio of same sign muon decays is
\begin{align}
R&\simeq\frac12\left[1-\frac{\left(1-r^2\right)^2}{\left(1+x^2\right)^2\left(1+r^2\right)^2-4x^2r^2\sin^2\beta}\right].
\end{align}
For $x\gtrsim r$, as we would expect without fine-tuning, the product of the asymmetry and the fraction of same sign decays is approximately
\begin{align}
A\times R&\simeq\frac{2xr\sin\beta}{\left(x^2+1\right)^2}.
\end{align}

In the benchmark model we will consider in Sec.~\ref{sec:benchmark}, the final states involving muons common to both $R=\pm1$ states  dominate the total width, which corresponds to $\Gamma_0\ll\Gamma$ in Eq.~(\ref{eq:pdgham}), and we can no longer ignore the width difference or the deviation of $|p/q|$ from unity. In this case, $\Delta\Gamma$, $|p/q|-1\propto r$. For $r<2M_M/\Gamma$, which we expect is the case in the absence of fine-tuning, we can write the asymmetry as
\begin{align}\label{eq:asymmapprox}
A&\simeq\frac{4r}{x}\left(\frac{x^2+3}{x^2+2}\right)\sin\beta,
\end{align}
and the fraction of same sign decays as
\begin{align}\label{eq:fractionapprox}
R&\simeq\frac{x^2}{2}\frac{x^2+2}{\left(x^2+1\right)^2}.
\end{align}

The asymmetries that we have expressed above can be significant for a fairly wide range of parameters, and the product of the asymmetry and the fraction of same sign decays is typically of order $x\,r\sin\beta$. We also note that when $r$ is close to one, $A$ is suppressed for any value of $\Delta\Gamma$ and $|p/q|$.

\section{Benchmark Model Estimates}
\label{sec:benchmark}
Here  we give sample parameters which allow for sizable \emph{CP} violation in gluino decays. The distinctive final state that the gluino decays into through $\call_{\rm eff}$ in Eq.~(\ref{eq:Leff2}), $\mu jj$ is subject to leptoquark searches at the LHC~\cite{Evans:2013uwa}.\footnote{This scenario could lead to a $\ell\ell+{\rm jets}$ signal for $\ell=e,\,\mu,\,\tau$, an intriguing possibility in light of recent excesses in leptoquark~\cite{CMS:2014qpa} and right-handed charged gauge boson~\cite{Khachatryan:2014dka} searches.} The very strong constraints from CMS on second generation leptoquarks using $20~{\rm fb}^{-1}$ of 8~TeV data~\cite{CMS-PAS-EXO-12-042} suggest that a gluino that decays with an $O(1)$ branching fraction to this final state should be heavy enough to be out of reach at 8~TeV. We therefore choose a benchmark gluino mass of 1.6~TeV, out of the reach of this search as well as standard SUSY searches involving missing energy. At next-to-leading order in QCD including next-to-leading-logarithmic threshold corrections, assuming the squarks are decoupled, the cross section for a 1.6~TeV Dirac gluino pair production in $pp$ collisions is $16$~fb ($0.4$~fb) at 13~TeV (8~TeV) center-of-mass energy, with an uncertainty on the order of 15-20\%~\cite{Beenakker:1996ch,*Kulesza:2008jb,*Kulesza:2009kq,*Beenakker:2009ha,*Beenakker:2011fu}, which is in agreement with the limit from~\cite{CMS-PAS-EXO-12-042} given a 100\% branching to $\mu jj$.

The following estimate shows that we do not expect an observably long lifetime for the gluino unless $x\gg 1$, in which case \emph{CP} violation from interference between mixing and decay becomes suppressed. The mass splitting from anomaly mediation is proportional to the gravitino mass, while the rate for decay into a gluon and gravitino is  inversely proportional to the square of the gravitino mass.   We  cannot take the mass splitting to be small without taking  the gravitino  light or fine-tuning, however if we take the gravitino mass to be too small  the gluino will decay too fast to oscillate.  The rate for a gluino of mass $M_D$ to decay to a gluon and gravitino is~\cite{Goodsell:2014dia} 
\begin{equation}
\Gamma_{g \tilde{G}}=\frac{M_D^5}{12 M_{\rm Pl}^2 m_{3/2}^2} \label{eq:rate}
\end{equation} 
which gives  $\Gamma_{g \tilde{G}}\sim 60~$eV for a 1.6~TeV gluino mass and 10~eV gravitino. From Eq.~(\ref{eq:mlambda}), a gravitino mass of 10~eV would give a mass splitting from anomaly mediation of about 0.4~eV. We therefore can only have comparable oscillation and decay rates when the gravitino is heavier than a few eV and the gluino width is greater than about an eV.  

We consider a gluino width of 300~eV and assume the gravitino branching fraction is small, so that the decays are dominated by the effective operators of Eq.~(\ref{eq:Leff2}).    For a 1.6~TeV gluino this width corresponds to  
\begin{equation}\sqrt{|{\tilde{G}}_\lambda|^2 +|{\tilde{G}}_\calo|^2}\sim \frac{1}{(21~{\rm TeV})^2}.\end{equation} Taking $\mu_D=5$~TeV, $m_{\phi_{\bar{d}}}=4$~TeV, $|g'_1|=|y^\prime_{21}| = 0.3$, $y_{211}=0.02$, and $|B^2_{D\bar{D}}|=(1.25\ {\rm TeV})^2$ gives
\begin{align}|\tilde{G}_\lambda|=&\frac{1}{(21 \ {\rm TeV})^2},~|\tilde{G}_\calo|= \frac{1}{(56 \ {\rm TeV})^2}\ .\end{align} Given  scalar masses of this size, these values of the $R$-parity--violating couplings $y_{211}$ and $y^\prime_{21}$ are in agreement with limits from charged pion decays and neutrino scattering~\cite{Barbier:2004ez}. The ratio $r$ is
\begin{equation}
r=\frac{|\calm_+^+|}{|\calm_-^+|}=\frac{|\tilde{G}_\calo|}{|\tilde{G}_\lambda|}=0.14.
\end{equation}
We work in a basis where we have rotated the Majorana mass to be real. The phase $\phi_\Gamma\equiv\arg\Gamma_{12}$ is a free parameter and is related to the physical phase in this basis as $\phi_\Gamma\simeq\beta+\pi$. Loop corrections to the $U(1)_R$-breaking gaugino mass splitting are effectively at the two loop level, due to our assumption of a flavor symmetry suppressing $\tilde m_k$ and $B_{D\bar d_k}$, and are of order $r\Gamma$. This means that without fine-tuning, $M_M\simeq x\Gamma/2\gtrsim r\Gamma$. The particular value, however, depends on the gravitino mass and is a free parameter. Taking the mass splitting to be 200~eV and $\phi_\Gamma=-\pi/3$ gives a dimuon asymmetry
\begin{equation}
A\simeq 0.8,
\end{equation}
and a fraction of same sign events of $R\simeq0.25$. Given the production cross sections above, we therefore expect about $400$ (2) same sign muon pair events in $100~{\rm fb}^{-1}$ of data at 13~TeV ($20~{\rm fb}^{-1}$ at 8~TeV). This event rate could allow for $O(10\%)$ asymmetries to be probed.

\begin{figure}
\includegraphics[width=\linewidth]{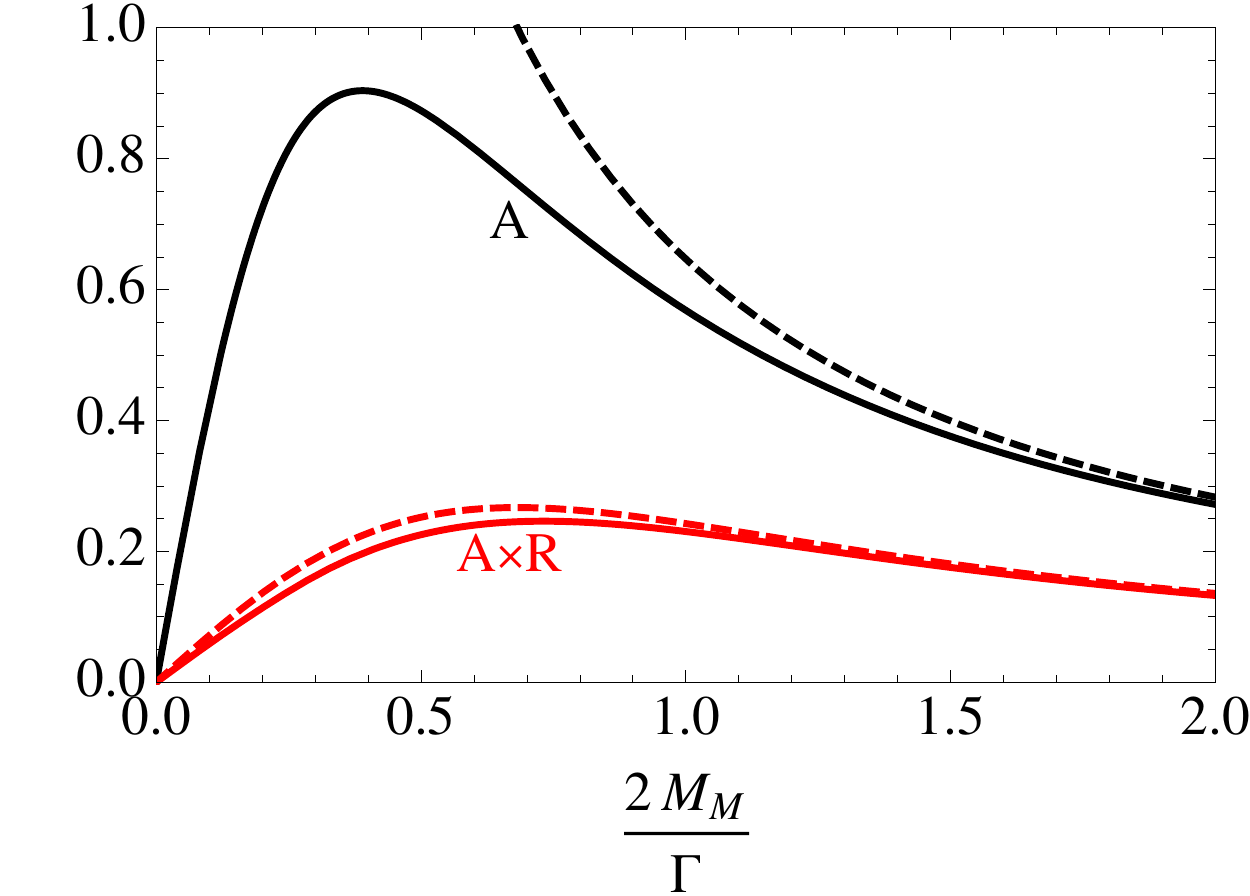}
\caption{The same sign dimuon asymmetry, $A$, of pseudo Dirac gluino decays as defined in Eq.~(\ref{eq:asymmetry}) (upper/black solid curve) and the approximate expression for $A$ in Eq.~(\ref{eq:asymmapprox}) (upper/black dashed curve) along with the product of $A$ and the ratio of same sign muon decays, $R$, defined in Eq.~(\ref{eq:fraction}) (lower/red solid curve) and the product of the approximate expressions for $A$ [Eq.~(\ref{eq:asymmapprox})] and $R$ [Eq.~(\ref{eq:fractionapprox})] (lower/red dashed curve) as functions of $2M_M/\Gamma\simeq x=\Delta m/\Gamma$. We have taken $\Gamma=300~{\rm eV}$, $r=0.14$, and $\phi_\Gamma=-\pi/3$.}\label{fig:asymm}
\end{figure}

In Fig.~\ref{fig:asymm}, we show the asymmetry for the parameters specified above, allowing the mass splitting to vary, as well as the approximate expression for the asymmetry from Eq.~(\ref{eq:asymmapprox}). We also show the product of the asymmetry and the ratio of same sign decays and the product of the approximate expressions in Eqs.~(\ref{eq:asymmapprox}) and (\ref{eq:fractionapprox}).

Note that assuming this gluino mass splitting is dominated by the anomaly-mediated contribution gives a gravitino mass of about 5~keV,  which could make the gravitino an interesting warm dark matter candidate. A 5~keV gravitino mass gives a branching fraction for the gluino to gluon plus gravitino of   $0.8\times 10^{-6}$.

\section{Summary and Outlook} \label{sec:conc}

This paper is the first to study the possibility of \emph{CP} violation in the decays of oscillating pseudo-Dirac fermions. We set up the effective  Hamiltonian, and show that it takes the same form as the one used for decays of oscillating mesons. We then consider a particular example, chosen to have the distinctive signature of an asymmetry between pairs of positively and negatively charged muons produced from gluino decays.   Similar phenomena are possible for a pseudo-Dirac neutralino. We note that order one asymmetries  in like sign dilepton events are possible.

Another possibility for heavy decaying pseudo-Dirac fermions is a supersymmetric theory (not necessarily containing  an approximate $U(1)_R$ symmetry or pseudo-Dirac gauginos)  with squarks as the lightest superpartners, in which case the squarks may hadronize as mesinos before they decay via R-parity violation. \emph{CP} violation from interference between oscillation and decays would be a generic feature of mesino decays as well.

Besides the unusual signature, our example was motivated by the $U(1)_R$ symmetry solution to the SUSY \emph{CP} problem, and the potential to obtain large \emph{CP} violation for baryogenesis which is not constrained by electric dipole moments. If the  lightest particle of the MSSM (besides the gravitino) is a pseudo-Dirac fermion which decays primarily via R-parity violation, \emph{CP} violation in the decays could produce either a baryon asymmetry or a lepton asymmetry which gets converted by anomalous weak processes into a baryon asymmetry. If such a particle could also be produced in a collider, then the \emph{CP} violation responsible for baryogenesis could potentially be directly observed. 

\begin{acknowledgments}
We thank Sean Tulin and Yuhsin Tsai for helpful discussions. We also thank Mark Goodsell for pointing out a factor of two in Eq.~(\ref{eq:rate}) in an earlier version of this manuscript. This work was supported in part by the U.S. Department of Energy under Grant No. DE-FG02-96ER40956. DM also acknowledges the Aspen Center for Physics, funded by NSF Grant No. 1066293, where part of this work was completed.
\end{acknowledgments}
\appendix
\section{Strong Interactions and Decoherence}
\label{sec:decohere}
Whether strong interactions decohere the color adjoint fermions can be analyzed 
by considering the time evolution of the density matrix (see 
Ref.~\cite{Tulin:2012re} for a detailed derivation and discussion of this 
formalism),
\begin{equation}
{\bm\rho}=\sum_{i,j=\psi,\bar \psi}|i\rangle\langle j|,
\end{equation}
which normally evolves in time as
\begin{equation}
\frac{\partial{\bm\rho}}{\partial t}=-i\left[{\bm H},{\bm\rho}\right],
\label{eq:dens_ev0}
\end{equation}
where ${\bm H}$ is the Hamiltonian. 
Including scattering off of sources of color charge (e.g. quarks, $\psi\,q\to \psi\,q$ and 
$\bar \psi\,q\to \bar \psi\,q$) modifies the evolution equation to
\begin{equation}
\frac{\partial{\bm\rho}}{\partial t}=-i\left[{\bm H},{\bm\rho}\right]-\frac{\kappa}{2}\left[{\bm N},\left[{\bm N},{\bm\rho}\right]\right].
\label{eq:dens_ev}
\end{equation}
$\kappa>0$ parameterizes the strength of the interaction 
and ${\bm N}$ is a matrix given 
by ${\bm N}={\rm diag}\left(1,\pm 1\right)$. The sign of the last term in ${\bm N}$ 
is determined by the transformation of the interaction Lagrangian under charge 
conjugation, $C$, of {\em only} the color adjoints in question,
$\psi\leftrightarrow\bar \psi$, $\cal L_{\rm int}\to\pm\cal L_{\rm int}$. If this is a 
minus sign, the interactions can distinguish between particle and antiparticle 
and the last term of Eq.~(\ref{eq:dens_ev}) becomes 
\begin{equation}
\left[{\bm N},\left[{\bm N},{\bm\rho}\right]\right]\propto\left(
  \begin{array}{cc}
    0 & \rho_{\psi\bar\psi} \\ 
    \rho_{\bar\psi\psi} & 0 
  \end{array}\right).
\end{equation}

This causes decoherence and can suppress oscillations. However, if the 
interactions cannot tell the difference between $\psi$ and $\bar \psi$ then ${\bm N}$ is 
the identity matrix so the last term in Eq.~(\ref{eq:dens_ev}) vanishes and 
coherent oscillations can occur.

In the case we consider, the interactions can be written simply as
\begin{equation}
{\cal L}_{\rm int}=ig_s\bar \psi T^a\gamma^\mu \psi J^a_\mu.
\end{equation}
where $T^a$ is a generator in the adjoint of SU(3), $J^a_\mu$ is a source of 
color charge, and $a=1,\dots,8$. Acting with $C$ on $\psi$ and $\bar \psi$ alone,
\begin{equation}
{\cal L}_{\rm int}\to-ig_s\bar \psi \left(T^a\right)^{\rm T}\gamma^\mu \psi J^a_\mu.
\end{equation}
Since $\psi$ and $\bar \psi$ are in the adjoint representation, $T^a$ is antisymmetric 
and $\cal L_{\rm int}\to\cal L_{\rm int}$. Thus, strong rescatterings do not 
decohere pseudo-Dirac gluinos and they can undergo oscillations in the same way
as (pseudo-Dirac) electroweak gauginos.

\bibliography{ref}

\begin{thebibliography}{64}%
\makeatletter
\providecommand \@ifxundefined [1]{%
 \@ifx{#1\undefined}
}%
\providecommand \@ifnum [1]{%
 \ifnum #1\expandafter \@firstoftwo
 \else \expandafter \@secondoftwo
 \fi
}%
\providecommand \@ifx [1]{%
 \ifx #1\expandafter \@firstoftwo
 \else \expandafter \@secondoftwo
 \fi
}%
\providecommand \natexlab [1]{#1}%
\providecommand \enquote  [1]{``#1''}%
\providecommand \bibnamefont  [1]{#1}%
\providecommand \bibfnamefont [1]{#1}%
\providecommand \citenamefont [1]{#1}%
\providecommand \href@noop [0]{\@secondoftwo}%
\providecommand \href [0]{\begingroup \@sanitize@url \@href}%
\providecommand \@href[1]{\@@startlink{#1}\@@href}%
\providecommand \@@href[1]{\endgroup#1\@@endlink}%
\providecommand \@sanitize@url [0]{\catcode `\\12\catcode `\$12\catcode
  `\&12\catcode `\#12\catcode `\^12\catcode `\_12\catcode `\%12\relax}%
\providecommand \@@startlink[1]{}%
\providecommand \@@endlink[0]{}%
\providecommand \url  [0]{\begingroup\@sanitize@url \@url }%
\providecommand \@url [1]{\endgroup\@href {#1}{\urlprefix }}%
\providecommand \urlprefix  [0]{URL }%
\providecommand \Eprint [0]{\href }%
\providecommand \doibase [0]{http://dx.doi.org/}%
\providecommand \selectlanguage [0]{\@gobble}%
\providecommand \bibinfo  [0]{\@secondoftwo}%
\providecommand \bibfield  [0]{\@secondoftwo}%
\providecommand \translation [1]{[#1]}%
\providecommand \BibitemOpen [0]{}%
\providecommand \bibitemStop [0]{}%
\providecommand \bibitemNoStop [0]{.\EOS\space}%
\providecommand \EOS [0]{\spacefactor3000\relax}%
\providecommand \BibitemShut  [1]{\csname bibitem#1\endcsname}%
\let\auto@bib@innerbib\@empty
\bibitem [{\citenamefont {Kuzmin}\ \emph {et~al.}(1985)\citenamefont {Kuzmin},
  \citenamefont {Rubakov},\ and\ \citenamefont {Shaposhnikov}}]{Kuzmin:1985mm}%
  \BibitemOpen
  \bibfield  {author} {\bibinfo {author} {\bibfnamefont {V.}~\bibnamefont
  {Kuzmin}}, \bibinfo {author} {\bibfnamefont {V.}~\bibnamefont {Rubakov}}, \
  and\ \bibinfo {author} {\bibfnamefont {M.}~\bibnamefont {Shaposhnikov}},\
  }\href {\doibase 10.1016/0370-2693(85)91028-7} {\bibfield  {journal}
  {\bibinfo  {journal} {Phys. Lett. B}\ }\textbf {\bibinfo {volume} {155}},\
  \bibinfo {pages} {36} (\bibinfo {year} {1985})}\BibitemShut {NoStop}%
\bibitem [{\citenamefont {Huet}\ and\ \citenamefont
  {Sather}(1995)}]{Huet:1994jb}%
  \BibitemOpen
  \bibfield  {author} {\bibinfo {author} {\bibfnamefont {P.}~\bibnamefont
  {Huet}}\ and\ \bibinfo {author} {\bibfnamefont {E.}~\bibnamefont {Sather}},\
  }\href {\doibase 10.1103/PhysRevD.51.379} {\bibfield  {journal} {\bibinfo
  {journal} {Phys. Rev. D}\ }\textbf {\bibinfo {volume} {51}},\ \bibinfo
  {pages} {379} (\bibinfo {year} {1995})},\ \Eprint
  {http://arxiv.org/abs/hep-ph/9404302} {arXiv:hep-ph/9404302 [hep-ph]}
  \BibitemShut {NoStop}%
\bibitem [{\citenamefont {Gavela}\ \emph {et~al.}(1994)\citenamefont {Gavela},
  \citenamefont {Hernandez}, \citenamefont {Orloff}, \citenamefont {Pene},\
  and\ \citenamefont {Quimbay}}]{Gavela:1994dt}%
  \BibitemOpen
  \bibfield  {author} {\bibinfo {author} {\bibfnamefont {M.}~\bibnamefont
  {Gavela}}, \bibinfo {author} {\bibfnamefont {P.}~\bibnamefont {Hernandez}},
  \bibinfo {author} {\bibfnamefont {J.}~\bibnamefont {Orloff}}, \bibinfo
  {author} {\bibfnamefont {O.}~\bibnamefont {Pene}}, \ and\ \bibinfo {author}
  {\bibfnamefont {C.}~\bibnamefont {Quimbay}},\ }\href {\doibase
  10.1016/0550-3213(94)00410-2} {\bibfield  {journal} {\bibinfo  {journal}
  {Nucl. Phys. B}\ }\textbf {\bibinfo {volume} {430}},\ \bibinfo {pages} {382}
  (\bibinfo {year} {1994})},\ \Eprint {http://arxiv.org/abs/hep-ph/9406289}
  {arXiv:hep-ph/9406289 [hep-ph]} \BibitemShut {NoStop}%
\bibitem [{\citenamefont {Pospelov}\ and\ \citenamefont
  {Ritz}(2005)}]{Pospelov:2005pr}%
  \BibitemOpen
  \bibfield  {author} {\bibinfo {author} {\bibfnamefont {M.}~\bibnamefont
  {Pospelov}}\ and\ \bibinfo {author} {\bibfnamefont {A.}~\bibnamefont
  {Ritz}},\ }\href {\doibase 10.1016/j.aop.2005.04.002} {\bibfield  {journal}
  {\bibinfo  {journal} {Annals Phys.}\ }\textbf {\bibinfo {volume} {318}},\
  \bibinfo {pages} {119} (\bibinfo {year} {2005})},\ \Eprint
  {http://arxiv.org/abs/hep-ph/0504231} {arXiv:hep-ph/0504231 [hep-ph]}
  \BibitemShut {NoStop}%
\bibitem [{\citenamefont {Cirigliano}\ \emph {et~al.}(2006)\citenamefont
  {Cirigliano}, \citenamefont {Profumo},\ and\ \citenamefont
  {Ramsey-Musolf}}]{Cirigliano:2006dg}%
  \BibitemOpen
  \bibfield  {author} {\bibinfo {author} {\bibfnamefont {V.}~\bibnamefont
  {Cirigliano}}, \bibinfo {author} {\bibfnamefont {S.}~\bibnamefont {Profumo}},
  \ and\ \bibinfo {author} {\bibfnamefont {M.~J.}\ \bibnamefont
  {Ramsey-Musolf}},\ }\href {\doibase 10.1088/1126-6708/2006/07/002} {\bibfield
   {journal} {\bibinfo  {journal} {JHEP}\ }\textbf {\bibinfo {volume} {0607}},\
  \bibinfo {pages} {002} (\bibinfo {year} {2006})},\ \Eprint
  {http://arxiv.org/abs/hep-ph/0603246} {arXiv:hep-ph/0603246 [hep-ph]}
  \BibitemShut {NoStop}%
\bibitem [{\citenamefont {Fromme}\ \emph {et~al.}(2006)\citenamefont {Fromme},
  \citenamefont {Huber},\ and\ \citenamefont {Seniuch}}]{Fromme:2006cm}%
  \BibitemOpen
  \bibfield  {author} {\bibinfo {author} {\bibfnamefont {L.}~\bibnamefont
  {Fromme}}, \bibinfo {author} {\bibfnamefont {S.~J.}\ \bibnamefont {Huber}}, \
  and\ \bibinfo {author} {\bibfnamefont {M.}~\bibnamefont {Seniuch}},\ }\href
  {\doibase 10.1088/1126-6708/2006/11/038} {\bibfield  {journal} {\bibinfo
  {journal} {JHEP}\ }\textbf {\bibinfo {volume} {0611}},\ \bibinfo {pages}
  {038} (\bibinfo {year} {2006})},\ \Eprint
  {http://arxiv.org/abs/hep-ph/0605242} {arXiv:hep-ph/0605242 [hep-ph]}
  \BibitemShut {NoStop}%
\bibitem [{\citenamefont {Morrissey}\ and\ \citenamefont
  {Ramsey-Musolf}(2012)}]{Morrissey:2012db}%
  \BibitemOpen
  \bibfield  {author} {\bibinfo {author} {\bibfnamefont {D.~E.}\ \bibnamefont
  {Morrissey}}\ and\ \bibinfo {author} {\bibfnamefont {M.~J.}\ \bibnamefont
  {Ramsey-Musolf}},\ }\href {\doibase 10.1088/1367-2630/14/12/125003}
  {\bibfield  {journal} {\bibinfo  {journal} {New J. Phys.}\ }\textbf {\bibinfo
  {volume} {14}},\ \bibinfo {pages} {125003} (\bibinfo {year} {2012})},\
  \Eprint {http://arxiv.org/abs/1206.2942} {arXiv:1206.2942 [hep-ph]}
  \BibitemShut {NoStop}%
\bibitem [{\citenamefont {McKeen}\ \emph {et~al.}(2012)\citenamefont {McKeen},
  \citenamefont {Pospelov},\ and\ \citenamefont {Ritz}}]{McKeen:2012av}%
  \BibitemOpen
  \bibfield  {author} {\bibinfo {author} {\bibfnamefont {D.}~\bibnamefont
  {McKeen}}, \bibinfo {author} {\bibfnamefont {M.}~\bibnamefont {Pospelov}}, \
  and\ \bibinfo {author} {\bibfnamefont {A.}~\bibnamefont {Ritz}},\ }\href
  {\doibase 10.1103/PhysRevD.86.113004} {\bibfield  {journal} {\bibinfo
  {journal} {Phys. Rev. D}\ }\textbf {\bibinfo {volume} {86}},\ \bibinfo
  {pages} {113004} (\bibinfo {year} {2012})},\ \Eprint
  {http://arxiv.org/abs/1208.4597} {arXiv:1208.4597 [hep-ph]} \BibitemShut
  {NoStop}%
\bibitem [{\citenamefont {McKeen}\ \emph {et~al.}(2013)\citenamefont {McKeen},
  \citenamefont {Pospelov},\ and\ \citenamefont {Ritz}}]{McKeen:2013dma}%
  \BibitemOpen
  \bibfield  {author} {\bibinfo {author} {\bibfnamefont {D.}~\bibnamefont
  {McKeen}}, \bibinfo {author} {\bibfnamefont {M.}~\bibnamefont {Pospelov}}, \
  and\ \bibinfo {author} {\bibfnamefont {A.}~\bibnamefont {Ritz}},\ }\href
  {\doibase 10.1103/PhysRevD.87.113002} {\bibfield  {journal} {\bibinfo
  {journal} {Phys. Rev. D}\ }\textbf {\bibinfo {volume} {87}},\ \bibinfo
  {pages} {113002} (\bibinfo {year} {2013})},\ \Eprint
  {http://arxiv.org/abs/1303.1172} {arXiv:1303.1172 [hep-ph]} \BibitemShut
  {NoStop}%
\bibitem [{\citenamefont {Ipek}(2014)}]{Ipek:2013iba}%
  \BibitemOpen
  \bibfield  {author} {\bibinfo {author} {\bibfnamefont {S.}~\bibnamefont
  {Ipek}},\ }\href {\doibase 10.1103/PhysRevD.89.073012} {\bibfield  {journal}
  {\bibinfo  {journal} {Phys. Rev. D}\ }\textbf {\bibinfo {volume} {89}},\
  \bibinfo {pages} {073012} (\bibinfo {year} {2014})},\ \Eprint
  {http://arxiv.org/abs/1310.6790} {arXiv:1310.6790 [hep-ph]} \BibitemShut
  {NoStop}%
\bibitem [{\citenamefont {Ellis}\ \emph {et~al.}(2008)\citenamefont {Ellis},
  \citenamefont {Lee},\ and\ \citenamefont {Pilaftsis}}]{Ellis:2008zy}%
  \BibitemOpen
  \bibfield  {author} {\bibinfo {author} {\bibfnamefont {J.~R.}\ \bibnamefont
  {Ellis}}, \bibinfo {author} {\bibfnamefont {J.~S.}\ \bibnamefont {Lee}}, \
  and\ \bibinfo {author} {\bibfnamefont {A.}~\bibnamefont {Pilaftsis}},\ }\href
  {\doibase 10.1088/1126-6708/2008/10/049} {\bibfield  {journal} {\bibinfo
  {journal} {JHEP}\ }\textbf {\bibinfo {volume} {0810}},\ \bibinfo {pages}
  {049} (\bibinfo {year} {2008})},\ \Eprint {http://arxiv.org/abs/0808.1819}
  {arXiv:0808.1819 [hep-ph]} \BibitemShut {NoStop}%
\bibitem [{\citenamefont {Baron}\ \emph {et~al.}(2014)\citenamefont {Baron}
  \emph {et~al.}}]{Baron:2013eja}%
  \BibitemOpen
  \bibfield  {author} {\bibinfo {author} {\bibfnamefont {J.}~\bibnamefont
  {Baron}} \emph {et~al.} (\bibinfo {collaboration} {ACME Collaboration}),\
  }\href {\doibase 10.1126/science.1248213} {\bibfield  {journal} {\bibinfo
  {journal} {Science}\ }\textbf {\bibinfo {volume} {343}},\ \bibinfo {pages}
  {269} (\bibinfo {year} {2014})},\ \Eprint {http://arxiv.org/abs/1310.7534}
  {arXiv:1310.7534 [physics.atom-ph]} \BibitemShut {NoStop}%
\bibitem [{\citenamefont {Griffith}\ \emph {et~al.}(2009)\citenamefont
  {Griffith}, \citenamefont {Swallows}, \citenamefont {Loftus}, \citenamefont
  {Romalis}, \citenamefont {Heckel} \emph {et~al.}}]{Griffith:2009zz}%
  \BibitemOpen
  \bibfield  {author} {\bibinfo {author} {\bibfnamefont {W.}~\bibnamefont
  {Griffith}}, \bibinfo {author} {\bibfnamefont {M.}~\bibnamefont {Swallows}},
  \bibinfo {author} {\bibfnamefont {T.}~\bibnamefont {Loftus}}, \bibinfo
  {author} {\bibfnamefont {M.}~\bibnamefont {Romalis}}, \bibinfo {author}
  {\bibfnamefont {B.}~\bibnamefont {Heckel}},  \emph {et~al.},\ }\href
  {\doibase 10.1103/PhysRevLett.102.101601} {\bibfield  {journal} {\bibinfo
  {journal} {Phys. Rev. Lett.}\ }\textbf {\bibinfo {volume} {102}},\ \bibinfo
  {pages} {101601} (\bibinfo {year} {2009})}\BibitemShut {NoStop}%
\bibitem [{\citenamefont {Baker}\ \emph {et~al.}(2006)\citenamefont {Baker},
  \citenamefont {Doyle}, \citenamefont {Geltenbort}, \citenamefont {Green},
  \citenamefont {van~der Grinten} \emph {et~al.}}]{Baker:2006ts}%
  \BibitemOpen
  \bibfield  {author} {\bibinfo {author} {\bibfnamefont {C.}~\bibnamefont
  {Baker}}, \bibinfo {author} {\bibfnamefont {D.}~\bibnamefont {Doyle}},
  \bibinfo {author} {\bibfnamefont {P.}~\bibnamefont {Geltenbort}}, \bibinfo
  {author} {\bibfnamefont {K.}~\bibnamefont {Green}}, \bibinfo {author}
  {\bibfnamefont {M.}~\bibnamefont {van~der Grinten}},  \emph {et~al.},\ }\href
  {\doibase 10.1103/PhysRevLett.97.131801} {\bibfield  {journal} {\bibinfo
  {journal} {Phys. Rev. Lett.}\ }\textbf {\bibinfo {volume} {97}},\ \bibinfo
  {pages} {131801} (\bibinfo {year} {2006})},\ \Eprint
  {http://arxiv.org/abs/hep-ex/0602020} {arXiv:hep-ex/0602020 [hep-ex]}
  \BibitemShut {NoStop}%
\bibitem [{\citenamefont {Hall}\ and\ \citenamefont
  {Randall}(1991)}]{Hall:1990hq}%
  \BibitemOpen
  \bibfield  {author} {\bibinfo {author} {\bibfnamefont {L.}~\bibnamefont
  {Hall}}\ and\ \bibinfo {author} {\bibfnamefont {L.}~\bibnamefont {Randall}},\
  }\href {\doibase 10.1016/0550-3213(91)90444-3} {\bibfield  {journal}
  {\bibinfo  {journal} {Nucl. Phys. B}\ }\textbf {\bibinfo {volume} {352}},\
  \bibinfo {pages} {289} (\bibinfo {year} {1991})}\BibitemShut {NoStop}%
\bibitem [{\citenamefont {Kribs}\ \emph {et~al.}(2008)\citenamefont {Kribs},
  \citenamefont {Poppitz},\ and\ \citenamefont {Weiner}}]{Kribs:2007ac}%
  \BibitemOpen
  \bibfield  {author} {\bibinfo {author} {\bibfnamefont {G.~D.}\ \bibnamefont
  {Kribs}}, \bibinfo {author} {\bibfnamefont {E.}~\bibnamefont {Poppitz}}, \
  and\ \bibinfo {author} {\bibfnamefont {N.}~\bibnamefont {Weiner}},\ }\href
  {\doibase 10.1103/PhysRevD.78.055010} {\bibfield  {journal} {\bibinfo
  {journal} {Phys. Rev. D}\ }\textbf {\bibinfo {volume} {78}},\ \bibinfo
  {pages} {055010} (\bibinfo {year} {2008})},\ \Eprint
  {http://arxiv.org/abs/0712.2039} {arXiv:0712.2039 [hep-ph]} \BibitemShut
  {NoStop}%
\bibitem [{\citenamefont {Dudas}\ \emph {et~al.}(2014)\citenamefont {Dudas},
  \citenamefont {Goodsell}, \citenamefont {Heurtier},\ and\ \citenamefont
  {Tziveloglou}}]{Dudas:2013gga}%
  \BibitemOpen
  \bibfield  {author} {\bibinfo {author} {\bibfnamefont {E.}~\bibnamefont
  {Dudas}}, \bibinfo {author} {\bibfnamefont {M.}~\bibnamefont {Goodsell}},
  \bibinfo {author} {\bibfnamefont {L.}~\bibnamefont {Heurtier}}, \ and\
  \bibinfo {author} {\bibfnamefont {P.}~\bibnamefont {Tziveloglou}},\ }\href
  {\doibase 10.1016/j.nuclphysb.2014.05.005} {\bibfield  {journal} {\bibinfo
  {journal} {Nucl. Phys. B}\ }\textbf {\bibinfo {volume} {884}},\ \bibinfo
  {pages} {632} (\bibinfo {year} {2014})},\ \Eprint
  {http://arxiv.org/abs/1312.2011} {arXiv:1312.2011 [hep-ph]} \BibitemShut
  {NoStop}%
\bibitem [{\citenamefont {Fayet}(1975)}]{Fayet:1974pd}%
  \BibitemOpen
  \bibfield  {author} {\bibinfo {author} {\bibfnamefont {P.}~\bibnamefont
  {Fayet}},\ }\href {\doibase 10.1016/0550-3213(75)90636-7} {\bibfield
  {journal} {\bibinfo  {journal} {Nucl. Phys. B}\ }\textbf {\bibinfo {volume}
  {90}},\ \bibinfo {pages} {104} (\bibinfo {year} {1975})}\BibitemShut
  {NoStop}%
\bibitem [{\citenamefont {Fayet}(1976)}]{Fayet:1975yi}%
  \BibitemOpen
  \bibfield  {author} {\bibinfo {author} {\bibfnamefont {P.}~\bibnamefont
  {Fayet}},\ }\href {\doibase 10.1016/0550-3213(76)90458-2} {\bibfield
  {journal} {\bibinfo  {journal} {Nucl. Phys. B}\ }\textbf {\bibinfo {volume}
  {113}},\ \bibinfo {pages} {135} (\bibinfo {year} {1976})}\BibitemShut
  {NoStop}%
\bibitem [{\citenamefont {Dine}\ and\ \citenamefont
  {MacIntire}(1992)}]{Dine:1992yw}%
  \BibitemOpen
  \bibfield  {author} {\bibinfo {author} {\bibfnamefont {M.}~\bibnamefont
  {Dine}}\ and\ \bibinfo {author} {\bibfnamefont {D.}~\bibnamefont
  {MacIntire}},\ }\href {\doibase 10.1103/PhysRevD.46.2594} {\bibfield
  {journal} {\bibinfo  {journal} {Phys. Rev. D}\ }\textbf {\bibinfo {volume}
  {46}},\ \bibinfo {pages} {2594} (\bibinfo {year} {1992})},\ \Eprint
  {http://arxiv.org/abs/hep-ph/9205227} {arXiv:hep-ph/9205227 [hep-ph]}
  \BibitemShut {NoStop}%
\bibitem [{\citenamefont {Fox}\ \emph {et~al.}(2002)\citenamefont {Fox},
  \citenamefont {Nelson},\ and\ \citenamefont {Weiner}}]{Fox:2002bu}%
  \BibitemOpen
  \bibfield  {author} {\bibinfo {author} {\bibfnamefont {P.~J.}\ \bibnamefont
  {Fox}}, \bibinfo {author} {\bibfnamefont {A.~E.}\ \bibnamefont {Nelson}}, \
  and\ \bibinfo {author} {\bibfnamefont {N.}~\bibnamefont {Weiner}},\ }\href
  {\doibase 10.1088/1126-6708/2002/08/035} {\bibfield  {journal} {\bibinfo
  {journal} {JHEP}\ }\textbf {\bibinfo {volume} {0208}},\ \bibinfo {pages}
  {035} (\bibinfo {year} {2002})},\ \Eprint
  {http://arxiv.org/abs/hep-ph/0206096} {arXiv:hep-ph/0206096 [hep-ph]}
  \BibitemShut {NoStop}%
\bibitem [{\citenamefont {Randall}\ and\ \citenamefont
  {Sundrum}(1999)}]{Randall:1998uk}%
  \BibitemOpen
  \bibfield  {author} {\bibinfo {author} {\bibfnamefont {L.}~\bibnamefont
  {Randall}}\ and\ \bibinfo {author} {\bibfnamefont {R.}~\bibnamefont
  {Sundrum}},\ }\href {\doibase 10.1016/S0550-3213(99)00359-4} {\bibfield
  {journal} {\bibinfo  {journal} {Nucl. Phys. B}\ }\textbf {\bibinfo {volume}
  {557}},\ \bibinfo {pages} {79} (\bibinfo {year} {1999})},\ \Eprint
  {http://arxiv.org/abs/hep-th/9810155} {arXiv:hep-th/9810155 [hep-th]}
  \BibitemShut {NoStop}%
\bibitem [{\citenamefont {Giudice}\ \emph {et~al.}(1998)\citenamefont
  {Giudice}, \citenamefont {Luty}, \citenamefont {Murayama},\ and\
  \citenamefont {Rattazzi}}]{Giudice:1998xp}%
  \BibitemOpen
  \bibfield  {author} {\bibinfo {author} {\bibfnamefont {G.~F.}\ \bibnamefont
  {Giudice}}, \bibinfo {author} {\bibfnamefont {M.~A.}\ \bibnamefont {Luty}},
  \bibinfo {author} {\bibfnamefont {H.}~\bibnamefont {Murayama}}, \ and\
  \bibinfo {author} {\bibfnamefont {R.}~\bibnamefont {Rattazzi}},\ }\href
  {\doibase 10.1088/1126-6708/1998/12/027} {\bibfield  {journal} {\bibinfo
  {journal} {JHEP}\ }\textbf {\bibinfo {volume} {9812}},\ \bibinfo {pages}
  {027} (\bibinfo {year} {1998})},\ \Eprint
  {http://arxiv.org/abs/hep-ph/9810442} {arXiv:hep-ph/9810442 [hep-ph]}
  \BibitemShut {NoStop}%
\bibitem [{\citenamefont {Grossman}\ \emph {et~al.}(2013)\citenamefont
  {Grossman}, \citenamefont {Shakya},\ and\ \citenamefont
  {Tsai}}]{Grossman:2012nn}%
  \BibitemOpen
  \bibfield  {author} {\bibinfo {author} {\bibfnamefont {Y.}~\bibnamefont
  {Grossman}}, \bibinfo {author} {\bibfnamefont {B.}~\bibnamefont {Shakya}}, \
  and\ \bibinfo {author} {\bibfnamefont {Y.}~\bibnamefont {Tsai}},\ }\href
  {\doibase 10.1103/PhysRevD.88.035026} {\bibfield  {journal} {\bibinfo
  {journal} {Phys. Rev. D}\ }\textbf {\bibinfo {volume} {88}},\ \bibinfo
  {pages} {035026} (\bibinfo {year} {2013})},\ \Eprint
  {http://arxiv.org/abs/1211.3121} {arXiv:1211.3121 [hep-ph]} \BibitemShut
  {NoStop}%
\bibitem [{\citenamefont {Sarid}\ and\ \citenamefont
  {Thomas}(2000)}]{Sarid:1999zx}%
  \BibitemOpen
  \bibfield  {author} {\bibinfo {author} {\bibfnamefont {U.}~\bibnamefont
  {Sarid}}\ and\ \bibinfo {author} {\bibfnamefont {S.~D.}\ \bibnamefont
  {Thomas}},\ }\href {\doibase 10.1103/PhysRevLett.85.1178} {\bibfield
  {journal} {\bibinfo  {journal} {Phys. Rev. Lett.}\ }\textbf {\bibinfo
  {volume} {85}},\ \bibinfo {pages} {1178} (\bibinfo {year} {2000})},\ \Eprint
  {http://arxiv.org/abs/hep-ph/9909349} {arXiv:hep-ph/9909349 [hep-ph]}
  \BibitemShut {NoStop}%
\bibitem [{\citenamefont {Berger}\ \emph {et~al.}(2013)\citenamefont {Berger},
  \citenamefont {Csaki}, \citenamefont {Grossman},\ and\ \citenamefont
  {Heidenreich}}]{Berger:2012mm}%
  \BibitemOpen
  \bibfield  {author} {\bibinfo {author} {\bibfnamefont {J.}~\bibnamefont
  {Berger}}, \bibinfo {author} {\bibfnamefont {C.}~\bibnamefont {Csaki}},
  \bibinfo {author} {\bibfnamefont {Y.}~\bibnamefont {Grossman}}, \ and\
  \bibinfo {author} {\bibfnamefont {B.}~\bibnamefont {Heidenreich}},\ }\href
  {\doibase 10.1140/epjc/s10052-013-2408-8} {\bibfield  {journal} {\bibinfo
  {journal} {Eur. Phys. J. C}\ }\textbf {\bibinfo {volume} {73}},\ \bibinfo
  {pages} {2408} (\bibinfo {year} {2013})},\ \Eprint
  {http://arxiv.org/abs/1209.4645} {arXiv:1209.4645 [hep-ph]} \BibitemShut
  {NoStop}%
\bibitem [{\citenamefont {Bilenky}\ and\ \citenamefont
  {Petcov}(1987)}]{RevModPhys.59.671}%
  \BibitemOpen
  \bibfield  {author} {\bibinfo {author} {\bibfnamefont {S.~M.}\ \bibnamefont
  {Bilenky}}\ and\ \bibinfo {author} {\bibfnamefont {S.~T.}\ \bibnamefont
  {Petcov}},\ }\href {\doibase 10.1103/RevModPhys.59.671} {\bibfield  {journal}
  {\bibinfo  {journal} {Rev. Mod. Phys.}\ }\textbf {\bibinfo {volume} {59}},\
  \bibinfo {pages} {671} (\bibinfo {year} {1987})}\BibitemShut {NoStop}%
\bibitem [{\citenamefont {Wolfenstein}(1981)}]{Wolfenstein1981147}%
  \BibitemOpen
  \bibfield  {author} {\bibinfo {author} {\bibfnamefont {L.}~\bibnamefont
  {Wolfenstein}},\ }\href {\doibase
  http://dx.doi.org/10.1016/0550-3213(81)90096-1} {\bibfield  {journal}
  {\bibinfo  {journal} {Nuclear Physics B}\ }\textbf {\bibinfo {volume}
  {186}},\ \bibinfo {pages} {147 } (\bibinfo {year} {1981})}\BibitemShut
  {NoStop}%
\bibitem [{\citenamefont {Petcov}(1982)}]{Petcov1982245}%
  \BibitemOpen
  \bibfield  {author} {\bibinfo {author} {\bibfnamefont {S.}~\bibnamefont
  {Petcov}},\ }\href {\doibase http://dx.doi.org/10.1016/0370-2693(82)91246-1}
  {\bibfield  {journal} {\bibinfo  {journal} {Physics Letters B}\ }\textbf
  {\bibinfo {volume} {110}},\ \bibinfo {pages} {245 } (\bibinfo {year}
  {1982})}\BibitemShut {NoStop}%
\bibitem [{\citenamefont {Kobayashi}\ and\ \citenamefont
  {Lim}(2001)}]{Kobayashi:2000md}%
  \BibitemOpen
  \bibfield  {author} {\bibinfo {author} {\bibfnamefont {M.}~\bibnamefont
  {Kobayashi}}\ and\ \bibinfo {author} {\bibfnamefont {C.}~\bibnamefont
  {Lim}},\ }\href {\doibase 10.1103/PhysRevD.64.013003} {\bibfield  {journal}
  {\bibinfo  {journal} {Phys. Rev. D}\ }\textbf {\bibinfo {volume} {64}},\
  \bibinfo {pages} {013003} (\bibinfo {year} {2001})},\ \Eprint
  {http://arxiv.org/abs/hep-ph/0012266} {arXiv:hep-ph/0012266 [hep-ph]}
  \BibitemShut {NoStop}%
\bibitem [{\citenamefont {Schechter}\ and\ \citenamefont
  {Valle}(1980)}]{PhysRevD.22.2227}%
  \BibitemOpen
  \bibfield  {author} {\bibinfo {author} {\bibfnamefont {J.}~\bibnamefont
  {Schechter}}\ and\ \bibinfo {author} {\bibfnamefont {J.~W.~F.}\ \bibnamefont
  {Valle}},\ }\href {\doibase 10.1103/PhysRevD.22.2227} {\bibfield  {journal}
  {\bibinfo  {journal} {Phys. Rev. D}\ }\textbf {\bibinfo {volume} {22}},\
  \bibinfo {pages} {2227} (\bibinfo {year} {1980})}\BibitemShut {NoStop}%
\bibitem [{\citenamefont {Bray}\ \emph {et~al.}(2007)\citenamefont {Bray},
  \citenamefont {Lee},\ and\ \citenamefont {Pilaftsis}}]{Bray:2007ru}%
  \BibitemOpen
  \bibfield  {author} {\bibinfo {author} {\bibfnamefont {S.}~\bibnamefont
  {Bray}}, \bibinfo {author} {\bibfnamefont {J.~S.}\ \bibnamefont {Lee}}, \
  and\ \bibinfo {author} {\bibfnamefont {A.}~\bibnamefont {Pilaftsis}},\ }\href
  {\doibase 10.1016/j.nuclphysb.2007.07.002} {\bibfield  {journal} {\bibinfo
  {journal} {Nucl. Phys. B}\ }\textbf {\bibinfo {volume} {786}},\ \bibinfo
  {pages} {95} (\bibinfo {year} {2007})},\ \Eprint
  {http://arxiv.org/abs/hep-ph/0702294} {arXiv:hep-ph/0702294 [HEP-PH]}
  \BibitemShut {NoStop}%
\bibitem [{\citenamefont {Dreiner}\ \emph {et~al.}(2010)\citenamefont
  {Dreiner}, \citenamefont {Haber},\ and\ \citenamefont
  {Martin}}]{Dreiner:2010}%
  \BibitemOpen
  \bibfield  {author} {\bibinfo {author} {\bibfnamefont {H.~K.}\ \bibnamefont
  {Dreiner}}, \bibinfo {author} {\bibfnamefont {H.~E.}\ \bibnamefont {Haber}},
  \ and\ \bibinfo {author} {\bibfnamefont {S.~P.}\ \bibnamefont {Martin}},\
  }\href {\doibase 10.1016/j.physrep.2010.05.002} {\bibfield  {journal}
  {\bibinfo  {journal} {Physics Reports}\ }\textbf {\bibinfo {volume} {494}},\
  \bibinfo {pages} {1} (\bibinfo {year} {2010})}\BibitemShut {NoStop}%
\bibitem [{\citenamefont {Beringer}\ \emph {et~al.}(2012)\citenamefont
  {Beringer} \emph {et~al.}}]{Beringer:1900zz}%
  \BibitemOpen
  \bibfield  {author} {\bibinfo {author} {\bibfnamefont {J.}~\bibnamefont
  {Beringer}} \emph {et~al.} (\bibinfo {collaboration} {Particle Data Group}),\
  }\href {\doibase 10.1103/PhysRevD.86.010001} {\bibfield  {journal} {\bibinfo
  {journal} {Phys. Rev. D}\ }\textbf {\bibinfo {volume} {86}},\ \bibinfo
  {pages} {010001} (\bibinfo {year} {2012})}\BibitemShut {NoStop}%
\bibitem [{\citenamefont {Martone}\ and\ \citenamefont
  {Robinson}(2012)}]{Martone:2011kh}%
  \BibitemOpen
  \bibfield  {author} {\bibinfo {author} {\bibfnamefont {M.}~\bibnamefont
  {Martone}}\ and\ \bibinfo {author} {\bibfnamefont {D.~J.}\ \bibnamefont
  {Robinson}},\ }\href {\doibase 10.1103/PhysRevD.85.045006} {\bibfield
  {journal} {\bibinfo  {journal} {Phys. Rev. D}\ }\textbf {\bibinfo {volume}
  {85}},\ \bibinfo {pages} {045006} (\bibinfo {year} {2012})},\ \Eprint
  {http://arxiv.org/abs/1103.3486} {arXiv:1103.3486 [hep-ph]} \BibitemShut
  {NoStop}%
\bibitem [{\citenamefont {Pilaftsis}(1997)}]{Pilaftsis:1997dr}%
  \BibitemOpen
  \bibfield  {author} {\bibinfo {author} {\bibfnamefont {A.}~\bibnamefont
  {Pilaftsis}},\ }\href {\doibase 10.1016/S0550-3213(97)00469-0} {\bibfield
  {journal} {\bibinfo  {journal} {Nucl. Phys. B}\ }\textbf {\bibinfo {volume}
  {504}},\ \bibinfo {pages} {61} (\bibinfo {year} {1997})},\ \Eprint
  {http://arxiv.org/abs/hep-ph/9702393} {arXiv:hep-ph/9702393 [hep-ph]}
  \BibitemShut {NoStop}%
\bibitem [{\citenamefont {Frugiuele}\ \emph
  {et~al.}(2013{\natexlab{a}})\citenamefont {Frugiuele}, \citenamefont
  {Gregoire}, \citenamefont {Kumar},\ and\ \citenamefont
  {Ponton}}]{Frugiuele:2012kp}%
  \BibitemOpen
  \bibfield  {author} {\bibinfo {author} {\bibfnamefont {C.}~\bibnamefont
  {Frugiuele}}, \bibinfo {author} {\bibfnamefont {T.}~\bibnamefont {Gregoire}},
  \bibinfo {author} {\bibfnamefont {P.}~\bibnamefont {Kumar}}, \ and\ \bibinfo
  {author} {\bibfnamefont {E.}~\bibnamefont {Ponton}},\ }\href {\doibase
  10.1007/JHEP05(2013)012} {\bibfield  {journal} {\bibinfo  {journal} {JHEP}\
  }\textbf {\bibinfo {volume} {1305}},\ \bibinfo {pages} {012} (\bibinfo {year}
  {2013}{\natexlab{a}})},\ \Eprint {http://arxiv.org/abs/1210.5257}
  {arXiv:1210.5257 [hep-ph]} \BibitemShut {NoStop}%
\bibitem [{\citenamefont {Bertuzzo}\ \emph {et~al.}(2014)\citenamefont
  {Bertuzzo}, \citenamefont {Frugiuele}, \citenamefont {Gregoire},\ and\
  \citenamefont {Ponton}}]{Bertuzzo:2014bwa}%
  \BibitemOpen
  \bibfield  {author} {\bibinfo {author} {\bibfnamefont {E.}~\bibnamefont
  {Bertuzzo}}, \bibinfo {author} {\bibfnamefont {C.}~\bibnamefont {Frugiuele}},
  \bibinfo {author} {\bibfnamefont {T.}~\bibnamefont {Gregoire}}, \ and\
  \bibinfo {author} {\bibfnamefont {E.}~\bibnamefont {Ponton}},\ }\href@noop {}
  {\  (\bibinfo {year} {2014})},\ \Eprint {http://arxiv.org/abs/1402.5432}
  {arXiv:1402.5432 [hep-ph]} \BibitemShut {NoStop}%
\bibitem [{\citenamefont {Davies}\ \emph {et~al.}(2011)\citenamefont {Davies},
  \citenamefont {March-Russell},\ and\ \citenamefont
  {McCullough}}]{Davies:2011mp}%
  \BibitemOpen
  \bibfield  {author} {\bibinfo {author} {\bibfnamefont {R.}~\bibnamefont
  {Davies}}, \bibinfo {author} {\bibfnamefont {J.}~\bibnamefont
  {March-Russell}}, \ and\ \bibinfo {author} {\bibfnamefont {M.}~\bibnamefont
  {McCullough}},\ }\href {\doibase 10.1007/JHEP04(2011)108} {\bibfield
  {journal} {\bibinfo  {journal} {JHEP}\ }\textbf {\bibinfo {volume} {1104}},\
  \bibinfo {pages} {108} (\bibinfo {year} {2011})},\ \Eprint
  {http://arxiv.org/abs/1103.1647} {arXiv:1103.1647 [hep-ph]} \BibitemShut
  {NoStop}%
\bibitem [{\citenamefont {Frugiuele}\ and\ \citenamefont
  {Gregoire}(2012)}]{Frugiuele:2011mh}%
  \BibitemOpen
  \bibfield  {author} {\bibinfo {author} {\bibfnamefont {C.}~\bibnamefont
  {Frugiuele}}\ and\ \bibinfo {author} {\bibfnamefont {T.}~\bibnamefont
  {Gregoire}},\ }\href {\doibase 10.1103/PhysRevD.85.015016} {\bibfield
  {journal} {\bibinfo  {journal} {Phys. Rev. D}\ }\textbf {\bibinfo {volume}
  {85}},\ \bibinfo {pages} {015016} (\bibinfo {year} {2012})},\ \Eprint
  {http://arxiv.org/abs/1107.4634} {arXiv:1107.4634 [hep-ph]} \BibitemShut
  {NoStop}%
\bibitem [{\citenamefont {Frugiuele}\ \emph
  {et~al.}(2013{\natexlab{b}})\citenamefont {Frugiuele}, \citenamefont
  {Gregoire}, \citenamefont {Kumar},\ and\ \citenamefont
  {Ponton}}]{Frugiuele:2012pe}%
  \BibitemOpen
  \bibfield  {author} {\bibinfo {author} {\bibfnamefont {C.}~\bibnamefont
  {Frugiuele}}, \bibinfo {author} {\bibfnamefont {T.}~\bibnamefont {Gregoire}},
  \bibinfo {author} {\bibfnamefont {P.}~\bibnamefont {Kumar}}, \ and\ \bibinfo
  {author} {\bibfnamefont {E.}~\bibnamefont {Ponton}},\ }\href {\doibase
  10.1007/JHEP03(2013)156} {\bibfield  {journal} {\bibinfo  {journal} {JHEP}\
  }\textbf {\bibinfo {volume} {1303}},\ \bibinfo {pages} {156} (\bibinfo {year}
  {2013}{\natexlab{b}})},\ \Eprint {http://arxiv.org/abs/1210.0541}
  {arXiv:1210.0541 [hep-ph]} \BibitemShut {NoStop}%
\bibitem [{\citenamefont {Riva}\ \emph {et~al.}(2013)\citenamefont {Riva},
  \citenamefont {Biggio},\ and\ \citenamefont {Pomarol}}]{Riva:2012hz}%
  \BibitemOpen
  \bibfield  {author} {\bibinfo {author} {\bibfnamefont {F.}~\bibnamefont
  {Riva}}, \bibinfo {author} {\bibfnamefont {C.}~\bibnamefont {Biggio}}, \ and\
  \bibinfo {author} {\bibfnamefont {A.}~\bibnamefont {Pomarol}},\ }\href
  {\doibase 10.1007/JHEP02(2013)081} {\bibfield  {journal} {\bibinfo  {journal}
  {JHEP}\ }\textbf {\bibinfo {volume} {1302}},\ \bibinfo {pages} {081}
  (\bibinfo {year} {2013})},\ \Eprint {http://arxiv.org/abs/1211.4526}
  {arXiv:1211.4526 [hep-ph]} \BibitemShut {NoStop}%
\bibitem [{\citenamefont {Ellis}\ \emph {et~al.}(1985)\citenamefont {Ellis},
  \citenamefont {Gelmini}, \citenamefont {Jarlskog}, \citenamefont {Ross},\
  and\ \citenamefont {Valle}}]{Ellis:1984gi}%
  \BibitemOpen
  \bibfield  {author} {\bibinfo {author} {\bibfnamefont {J.~R.}\ \bibnamefont
  {Ellis}}, \bibinfo {author} {\bibfnamefont {G.}~\bibnamefont {Gelmini}},
  \bibinfo {author} {\bibfnamefont {C.}~\bibnamefont {Jarlskog}}, \bibinfo
  {author} {\bibfnamefont {G.~G.}\ \bibnamefont {Ross}}, \ and\ \bibinfo
  {author} {\bibfnamefont {J.}~\bibnamefont {Valle}},\ }\href {\doibase
  10.1016/0370-2693(85)90157-1} {\bibfield  {journal} {\bibinfo  {journal}
  {Phys. Lett. B}\ }\textbf {\bibinfo {volume} {150}},\ \bibinfo {pages} {142}
  (\bibinfo {year} {1985})}\BibitemShut {NoStop}%
\bibitem [{\citenamefont {Ross}\ and\ \citenamefont
  {Valle}(1985)}]{Ross:1984yg}%
  \BibitemOpen
  \bibfield  {author} {\bibinfo {author} {\bibfnamefont {G.~G.}\ \bibnamefont
  {Ross}}\ and\ \bibinfo {author} {\bibfnamefont {J.}~\bibnamefont {Valle}},\
  }\href {\doibase 10.1016/0370-2693(85)91658-2} {\bibfield  {journal}
  {\bibinfo  {journal} {Phys. Lett. B}\ }\textbf {\bibinfo {volume} {151}},\
  \bibinfo {pages} {375} (\bibinfo {year} {1985})}\BibitemShut {NoStop}%
\bibitem [{\citenamefont {Barbier}\ \emph {et~al.}(2005)\citenamefont
  {Barbier}, \citenamefont {Berat}, \citenamefont {Besancon}, \citenamefont
  {Chemtob}, \citenamefont {Deandrea} \emph {et~al.}}]{Barbier:2004ez}%
  \BibitemOpen
  \bibfield  {author} {\bibinfo {author} {\bibfnamefont {R.}~\bibnamefont
  {Barbier}}, \bibinfo {author} {\bibfnamefont {C.}~\bibnamefont {Berat}},
  \bibinfo {author} {\bibfnamefont {M.}~\bibnamefont {Besancon}}, \bibinfo
  {author} {\bibfnamefont {M.}~\bibnamefont {Chemtob}}, \bibinfo {author}
  {\bibfnamefont {A.}~\bibnamefont {Deandrea}},  \emph {et~al.},\ }\href
  {\doibase 10.1016/j.physrep.2005.08.006} {\bibfield  {journal} {\bibinfo
  {journal} {Phys. Rept.}\ }\textbf {\bibinfo {volume} {420}},\ \bibinfo
  {pages} {1} (\bibinfo {year} {2005})},\ \Eprint
  {http://arxiv.org/abs/hep-ph/0406039} {arXiv:hep-ph/0406039 [hep-ph]}
  \BibitemShut {NoStop}%
\bibitem [{\citenamefont {Fayet}(1984)}]{Fayet:1984jt}%
  \BibitemOpen
  \bibfield  {author} {\bibinfo {author} {\bibfnamefont {P.}~\bibnamefont
  {Fayet}},\ }\href {\doibase 10.1016/0370-2693(84)91195-X} {\bibfield
  {journal} {\bibinfo  {journal} {Phys. Lett. B}\ }\textbf {\bibinfo {volume}
  {142}},\ \bibinfo {pages} {263} (\bibinfo {year} {1984})}\BibitemShut
  {NoStop}%
\bibitem [{\citenamefont {Csaki}\ \emph {et~al.}(2014)\citenamefont {Csaki},
  \citenamefont {Goodman}, \citenamefont {Pavesi},\ and\ \citenamefont
  {Shirman}}]{Csaki:2013fla}%
  \BibitemOpen
  \bibfield  {author} {\bibinfo {author} {\bibfnamefont {C.}~\bibnamefont
  {Csaki}}, \bibinfo {author} {\bibfnamefont {J.}~\bibnamefont {Goodman}},
  \bibinfo {author} {\bibfnamefont {R.}~\bibnamefont {Pavesi}}, \ and\ \bibinfo
  {author} {\bibfnamefont {Y.}~\bibnamefont {Shirman}},\ }\href {\doibase
  10.1103/PhysRevD.89.055005} {\bibfield  {journal} {\bibinfo  {journal} {Phys.
  Rev. D}\ }\textbf {\bibinfo {volume} {89}},\ \bibinfo {pages} {055005}
  (\bibinfo {year} {2014})},\ \Eprint {http://arxiv.org/abs/1310.4504}
  {arXiv:1310.4504 [hep-ph]} \BibitemShut {NoStop}%
\bibitem [{\citenamefont {Benakli}\ and\ \citenamefont
  {Goodsell}(2009)}]{Benakli:2008pg}%
  \BibitemOpen
  \bibfield  {author} {\bibinfo {author} {\bibfnamefont {K.}~\bibnamefont
  {Benakli}}\ and\ \bibinfo {author} {\bibfnamefont {M.}~\bibnamefont
  {Goodsell}},\ }\href {\doibase 10.1016/j.nuclphysb.2009.03.002} {\bibfield
  {journal} {\bibinfo  {journal} {Nucl. Phys. B}\ }\textbf {\bibinfo {volume}
  {816}},\ \bibinfo {pages} {185} (\bibinfo {year} {2009})},\ \Eprint
  {http://arxiv.org/abs/0811.4409} {arXiv:0811.4409 [hep-ph]} \BibitemShut
  {NoStop}%
\bibitem [{\citenamefont {Benakli}\ and\ \citenamefont
  {Goodsell}(2010)}]{Benakli:2010gi}%
  \BibitemOpen
  \bibfield  {author} {\bibinfo {author} {\bibfnamefont {K.}~\bibnamefont
  {Benakli}}\ and\ \bibinfo {author} {\bibfnamefont {M.}~\bibnamefont
  {Goodsell}},\ }\href {\doibase 10.1016/j.nuclphysb.2010.06.018} {\bibfield
  {journal} {\bibinfo  {journal} {Nucl. Phys. B}\ }\textbf {\bibinfo {volume}
  {840}},\ \bibinfo {pages} {1} (\bibinfo {year} {2010})},\ \Eprint
  {http://arxiv.org/abs/1003.4957} {arXiv:1003.4957 [hep-ph]} \BibitemShut
  {NoStop}%
\bibitem [{\citenamefont {Barnett}\ \emph {et~al.}(1993)\citenamefont
  {Barnett}, \citenamefont {Gunion},\ and\ \citenamefont
  {Haber}}]{Barnett:1993ea}%
  \BibitemOpen
  \bibfield  {author} {\bibinfo {author} {\bibfnamefont {R.~M.}\ \bibnamefont
  {Barnett}}, \bibinfo {author} {\bibfnamefont {J.~F.}\ \bibnamefont {Gunion}},
  \ and\ \bibinfo {author} {\bibfnamefont {H.~E.}\ \bibnamefont {Haber}},\
  }\href {\doibase 10.1016/0370-2693(93)91623-U} {\bibfield  {journal}
  {\bibinfo  {journal} {Phys. Lett. B}\ }\textbf {\bibinfo {volume} {315}},\
  \bibinfo {pages} {349} (\bibinfo {year} {1993})},\ \Eprint
  {http://arxiv.org/abs/hep-ph/9306204} {arXiv:hep-ph/9306204 [hep-ph]}
  \BibitemShut {NoStop}%
\bibitem [{\citenamefont {Guchait}\ and\ \citenamefont
  {Roy}(1995)}]{Guchait:1994zk}%
  \BibitemOpen
  \bibfield  {author} {\bibinfo {author} {\bibfnamefont {M.}~\bibnamefont
  {Guchait}}\ and\ \bibinfo {author} {\bibfnamefont {D.}~\bibnamefont {Roy}},\
  }\href {\doibase 10.1103/PhysRevD.52.133} {\bibfield  {journal} {\bibinfo
  {journal} {Phys. Rev. D}\ }\textbf {\bibinfo {volume} {52}},\ \bibinfo
  {pages} {133} (\bibinfo {year} {1995})},\ \Eprint
  {http://arxiv.org/abs/hep-ph/9412329} {arXiv:hep-ph/9412329 [hep-ph]}
  \BibitemShut {NoStop}%
\bibitem [{\citenamefont {Baer}\ \emph {et~al.}(1992)\citenamefont {Baer},
  \citenamefont {Tata},\ and\ \citenamefont {Woodside}}]{Baer:1991xs}%
  \BibitemOpen
  \bibfield  {author} {\bibinfo {author} {\bibfnamefont {H.}~\bibnamefont
  {Baer}}, \bibinfo {author} {\bibfnamefont {X.}~\bibnamefont {Tata}}, \ and\
  \bibinfo {author} {\bibfnamefont {J.}~\bibnamefont {Woodside}},\ }\href
  {\doibase 10.1103/PhysRevD.45.142} {\bibfield  {journal} {\bibinfo  {journal}
  {Phys. Rev. D}\ }\textbf {\bibinfo {volume} {45}},\ \bibinfo {pages} {142}
  (\bibinfo {year} {1992})}\BibitemShut {NoStop}%
\bibitem [{\citenamefont {Baer}\ \emph {et~al.}(1996)\citenamefont {Baer},
  \citenamefont {Chen}, \citenamefont {Paige},\ and\ \citenamefont
  {Tata}}]{Baer:1995va}%
  \BibitemOpen
  \bibfield  {author} {\bibinfo {author} {\bibfnamefont {H.}~\bibnamefont
  {Baer}}, \bibinfo {author} {\bibfnamefont {C.-h.}\ \bibnamefont {Chen}},
  \bibinfo {author} {\bibfnamefont {F.}~\bibnamefont {Paige}}, \ and\ \bibinfo
  {author} {\bibfnamefont {X.}~\bibnamefont {Tata}},\ }\href {\doibase
  10.1103/PhysRevD.53.6241} {\bibfield  {journal} {\bibinfo  {journal} {Phys.
  Rev. D}\ }\textbf {\bibinfo {volume} {53}},\ \bibinfo {pages} {6241}
  (\bibinfo {year} {1996})},\ \Eprint {http://arxiv.org/abs/hep-ph/9512383}
  {arXiv:hep-ph/9512383 [hep-ph]} \BibitemShut {NoStop}%
\bibitem [{\citenamefont {Evans}\ and\ \citenamefont
  {Kats}(2013)}]{Evans:2013uwa}%
  \BibitemOpen
  \bibfield  {author} {\bibinfo {author} {\bibfnamefont {J.~A.}\ \bibnamefont
  {Evans}}\ and\ \bibinfo {author} {\bibfnamefont {Y.}~\bibnamefont {Kats}},\
  }\href@noop {} {\  (\bibinfo {year} {2013})},\ \Eprint
  {http://arxiv.org/abs/1311.0890} {arXiv:1311.0890 [hep-ph]} \BibitemShut
  {NoStop}%
\bibitem [{\citenamefont {{CMS Collaboration Public
  Note}}(2014)}]{CMS:2014qpa}%
  \BibitemOpen
  \bibfield  {author} {\bibinfo {author} {\bibnamefont {{CMS Collaboration
  Public Note}}},\ }\href@noop {} {\bibfield  {journal} {\bibinfo  {journal}
  {CMS-PAS-EXO-12-041,}\ } (\bibinfo {year} {2014})},\ \bibinfo {note}
  {\url{http://cds.cern.ch/record/1742179/files/EXO-12-041-pas.pdf}}\BibitemShut
  {NoStop}%
\bibitem [{\citenamefont {Khachatryan}\ \emph {et~al.}(2014)\citenamefont
  {Khachatryan} \emph {et~al.}}]{Khachatryan:2014dka}%
  \BibitemOpen
  \bibfield  {author} {\bibinfo {author} {\bibfnamefont {V.}~\bibnamefont
  {Khachatryan}} \emph {et~al.} (\bibinfo {collaboration} {CMS
  Collaboration}),\ }\href@noop {} {\  (\bibinfo {year} {2014})},\ \Eprint
  {http://arxiv.org/abs/1407.3683} {arXiv:1407.3683 [hep-ex]} \BibitemShut
  {NoStop}%
\bibitem [{\citenamefont {{CMS Collaboration Public
  Note}}(2013)}]{CMS-PAS-EXO-12-042}%
  \BibitemOpen
  \bibfield  {author} {\bibinfo {author} {\bibnamefont {{CMS Collaboration
  Public Note}}},\ }\href@noop {} {\bibfield  {journal} {\bibinfo  {journal}
  {CMS-PAS-EXO-12-042,}\ } (\bibinfo {year} {2013})},\ \bibinfo {note}
  {\url{http://cds.cern.ch/record/1542374/files/EXO-12-042-pas.pdf}}\BibitemShut
  {NoStop}%
\bibitem [{\citenamefont {Beenakker}\ \emph {et~al.}(1997)\citenamefont
  {Beenakker}, \citenamefont {Hopker}, \citenamefont {Spira},\ and\
  \citenamefont {Zerwas}}]{Beenakker:1996ch}%
  \BibitemOpen
  \bibfield  {author} {\bibinfo {author} {\bibfnamefont {W.}~\bibnamefont
  {Beenakker}}, \bibinfo {author} {\bibfnamefont {R.}~\bibnamefont {Hopker}},
  \bibinfo {author} {\bibfnamefont {M.}~\bibnamefont {Spira}}, \ and\ \bibinfo
  {author} {\bibfnamefont {P.}~\bibnamefont {Zerwas}},\ }\href {\doibase
  10.1016/S0550-3213(97)80027-2} {\bibfield  {journal} {\bibinfo  {journal}
  {Nucl. Phys. B}\ }\textbf {\bibinfo {volume} {492}},\ \bibinfo {pages} {51}
  (\bibinfo {year} {1997})},\ \Eprint {http://arxiv.org/abs/hep-ph/9610490}
  {arXiv:hep-ph/9610490 [hep-ph]} \BibitemShut {NoStop}%
\bibitem [{\citenamefont {Kulesza}\ and\ \citenamefont
  {Motyka}(2009{\natexlab{a}})}]{Kulesza:2008jb}%
  \BibitemOpen
  \bibfield  {author} {\bibinfo {author} {\bibfnamefont {A.}~\bibnamefont
  {Kulesza}}\ and\ \bibinfo {author} {\bibfnamefont {L.}~\bibnamefont
  {Motyka}},\ }\href {\doibase 10.1103/PhysRevLett.102.111802} {\bibfield
  {journal} {\bibinfo  {journal} {Phys. Rev. Lett.}\ }\textbf {\bibinfo
  {volume} {102}},\ \bibinfo {pages} {111802} (\bibinfo {year}
  {2009}{\natexlab{a}})},\ \Eprint {http://arxiv.org/abs/0807.2405}
  {arXiv:0807.2405 [hep-ph]} \BibitemShut {NoStop}%
\bibitem [{\citenamefont {Kulesza}\ and\ \citenamefont
  {Motyka}(2009{\natexlab{b}})}]{Kulesza:2009kq}%
  \BibitemOpen
  \bibfield  {author} {\bibinfo {author} {\bibfnamefont {A.}~\bibnamefont
  {Kulesza}}\ and\ \bibinfo {author} {\bibfnamefont {L.}~\bibnamefont
  {Motyka}},\ }\href {\doibase 10.1103/PhysRevD.80.095004} {\bibfield
  {journal} {\bibinfo  {journal} {Phys. Rev. D}\ }\textbf {\bibinfo {volume}
  {80}},\ \bibinfo {pages} {095004} (\bibinfo {year} {2009}{\natexlab{b}})},\
  \Eprint {http://arxiv.org/abs/0905.4749} {arXiv:0905.4749 [hep-ph]}
  \BibitemShut {NoStop}%
\bibitem [{\citenamefont {Beenakker}\ \emph {et~al.}(2009)\citenamefont
  {Beenakker}, \citenamefont {Brensing}, \citenamefont {Kramer}, \citenamefont
  {Kulesza}, \citenamefont {Laenen} \emph {et~al.}}]{Beenakker:2009ha}%
  \BibitemOpen
  \bibfield  {author} {\bibinfo {author} {\bibfnamefont {W.}~\bibnamefont
  {Beenakker}}, \bibinfo {author} {\bibfnamefont {S.}~\bibnamefont {Brensing}},
  \bibinfo {author} {\bibfnamefont {M.}~\bibnamefont {Kramer}}, \bibinfo
  {author} {\bibfnamefont {A.}~\bibnamefont {Kulesza}}, \bibinfo {author}
  {\bibfnamefont {E.}~\bibnamefont {Laenen}},  \emph {et~al.},\ }\href
  {\doibase 10.1088/1126-6708/2009/12/041} {\bibfield  {journal} {\bibinfo
  {journal} {JHEP}\ }\textbf {\bibinfo {volume} {0912}},\ \bibinfo {pages}
  {041} (\bibinfo {year} {2009})},\ \Eprint {http://arxiv.org/abs/0909.4418}
  {arXiv:0909.4418 [hep-ph]} \BibitemShut {NoStop}%
\bibitem [{\citenamefont {Beenakker}\ \emph {et~al.}(2011)\citenamefont
  {Beenakker}, \citenamefont {Brensing}, \citenamefont {Kramer}, \citenamefont
  {Kulesza}, \citenamefont {Laenen} \emph {et~al.}}]{Beenakker:2011fu}%
  \BibitemOpen
  \bibfield  {author} {\bibinfo {author} {\bibfnamefont {W.}~\bibnamefont
  {Beenakker}}, \bibinfo {author} {\bibfnamefont {S.}~\bibnamefont {Brensing}},
  \bibinfo {author} {\bibfnamefont {M.}~\bibnamefont {Kramer}}, \bibinfo
  {author} {\bibfnamefont {A.}~\bibnamefont {Kulesza}}, \bibinfo {author}
  {\bibfnamefont {E.}~\bibnamefont {Laenen}},  \emph {et~al.},\ }\href
  {\doibase 10.1142/S0217751X11053560} {\bibfield  {journal} {\bibinfo
  {journal} {Int. J. Mod. Phys. A}\ }\textbf {\bibinfo {volume} {26}},\
  \bibinfo {pages} {2637} (\bibinfo {year} {2011})},\ \Eprint
  {http://arxiv.org/abs/1105.1110} {arXiv:1105.1110 [hep-ph]} \BibitemShut
  {NoStop}%
\bibitem [{\citenamefont {Goodsell}\ and\ \citenamefont
  {Tziveloglou}(2014)}]{Goodsell:2014dia}%
  \BibitemOpen
  \bibfield  {author} {\bibinfo {author} {\bibfnamefont {M.~D.}\ \bibnamefont
  {Goodsell}}\ and\ \bibinfo {author} {\bibfnamefont {P.}~\bibnamefont
  {Tziveloglou}},\ }\href@noop {} {\  (\bibinfo {year} {2014})},\ \Eprint
  {http://arxiv.org/abs/1407.5076} {arXiv:1407.5076 [hep-ph]} \BibitemShut
  {NoStop}%
\bibitem [{\citenamefont {Tulin}\ \emph {et~al.}(2012)\citenamefont {Tulin},
  \citenamefont {Yu},\ and\ \citenamefont {Zurek}}]{Tulin:2012re}%
  \BibitemOpen
  \bibfield  {author} {\bibinfo {author} {\bibfnamefont {S.}~\bibnamefont
  {Tulin}}, \bibinfo {author} {\bibfnamefont {H.-B.}\ \bibnamefont {Yu}}, \
  and\ \bibinfo {author} {\bibfnamefont {K.~M.}\ \bibnamefont {Zurek}},\ }\href
  {\doibase 10.1088/1475-7516/2012/05/013} {\bibfield  {journal} {\bibinfo
  {journal} {JCAP}\ }\textbf {\bibinfo {volume} {1205}},\ \bibinfo {pages}
  {013} (\bibinfo {year} {2012})},\ \Eprint {http://arxiv.org/abs/1202.0283}
  {arXiv:1202.0283 [hep-ph]} \BibitemShut {NoStop}%
\end{thebibliography}%

\end{document}